%% file: main.tex
\documentclass[journal,10pt]{IEEEtran}
\IEEEoverridecommandlockouts

\newtheorem{definition}{Definition}
\newtheorem{theorem}{Theorem}
\newtheorem{lemma}[theorem]{Lemma}

\usepackage{hyperref}
\usepackage{booktabs}
\usepackage[caption=false]{subfig}
\usepackage{amsmath}
\usepackage{bm}
\usepackage{cite}
\usepackage{threeparttable}
\usepackage{pdfpages}
\usepackage{amsfonts}
\usepackage[T1]{fontenc}
\usepackage[utf8]{inputenc}
\newcommand{\mat}[1]{\bm{{#1}}}

\title{Optimal Recovery for Causal Inference\thanks{The authors are with the Coordinated Science Laboratory, University of Illinois Urbana-Champaign, Urbana, IL  61801, USA (e-mail: \{iferwna2, varshney\}@illinois.edu). 
 IF is also with the Department of Computer Science and LRV is also with the Department of Electrical and Computer Engineering.}\thanks{This work was supported in part by NSF grant ECCS-2033900 and the Center for Pathogen Diagnostics through the ZJU-UIUC Dynamic Engineering Science Interdisciplinary Research Enterprise (DESIRE).}} 
\author{Ibtihal Ferwana and
Lav R.\ Varshney, \IEEEmembership{IEEE Senior Member}}
\begin{document}

\maketitle

\begin{abstract}
Problems in causal inference can be fruitfully addressed using signal processing techniques.  As an example, it is crucial to successfully quantify the causal effects of an intervention to determine whether the intervention achieved desired outcomes. We present a new geometric signal processing approach to classical synthetic control called \emph{ellipsoidal optimal recovery (EOpR)}, for estimating the unobservable outcome of a treatment unit. EOpR provides policy evaluators with both worst-case and typical outcomes to help in decision making. It is an approximation-theoretic technique that relates to the theory of principal components, which recovers unknown observations given a learned signal class and a set of known observations. We show EOpR can improve pre-treatment fit and mitigate bias of the post-treatment estimate relative to other methods in causal inference. Beyond recovery of the unit of interest, an advantage of EOpR is that it produces worst-case limits over the estimates produced. We assess our approach on artificially-generated data, on datasets commonly used in the econometrics literature, and in the context of the COVID-19 pandemic, showing  better performance than baseline techniques. 
\end{abstract}

\begin{IEEEkeywords}
Optimal recovery, signal processing, causal inference, synthetic control 
\end{IEEEkeywords}

\section{Introduction}

A major component of policy evaluation is to estimate the effects of an implemented policy, so as to know whether it achieved its goals. Estimating effects further yield inferences of causal relationships between interventions and their outcomes. Quantifying the effect of a treatment has been a problem of interest not only in policy making but also across different domains in health sciences, social sciences, and engineering. Typically, the effect is measured by looking at the difference between outcomes before and after an intervention of a treated unit. However, for a given object (e.g., region) at a given time, only one of the outcomes is observed and not both. Thus, we aim to recover and estimate the outcome that was not observed. 

Several causal inference methods have been developed for observational studies to estimate the unobserved outcomes for a given intervention. For example \textit{synthetic control} (SC) \cite{abadie2021synthetic} constructs a weighted average of control units to act as a synthetic control unit to compare with the treated unit. Recently, there has been a growing literature that approaches causal inference from a matrix completion perspective. Proposals include approximating the control unit matrix using nuclear-norm minimization \cite{athey2021matrix}, using singular value decomposition \cite{Amjad18}, or by finding nearest neighbors \cite{agarwal2021causal_matrix_completion} for missing entries of a matrix to best match control units and the treated unit of interest.

A main limitation in previous work is that when there is insufficient data, especially data from only a small period of time, methods are unable to recover the true estimate \cite{Amjad2019_mRSC}. Further, the insufficiency or the low quality of data tend to create a poor pre-treatment fit which is a main source of bias in estimates in SC \cite{abadie2021synthetic}. 

To address this, we propose an approximation-theoretic approach from signal processing called \textit{ellipsoidal optimal recovery} (EOpR) based on minimizing the worst-case error, which guarantees an exact fit of the pre-intervention period and optimally recovers the unobserved outcome with reduced bias in the effect estimate. Further, under the assumptions underlying EOpR, we derive worst-case estimates of effects, which are very useful in policy evaluation. Especially in situations of uncertainty, it is important to acknowledge the most severe possible outcomes that could occur for a given policy. Determining a policy alternative is derived by possible good and bad outcomes it yields. EOpR provides policy evaluators with ``worst-case'' outcomes and ``typical'' outcomes to help in decision making. 

To be specific, within the potential outcomes framework of Rubin \cite{rubin1974causaleffect}, we consider a variation of the optimal recovery \cite{muresaP2004} algorithm to recover missing outcomes for causal inference and to obtain worst-case estimates of the causal effect. Unlike statistical approaches\cite{abadie2003economic, abadie2010synthetic} or low-rank matrix decomposition approaches \cite{Amjad18, Amjad2019_mRSC, athey2021matrix}, optimal recovery is expected to be robust \cite{kassam1985robust} given that it minimizes the maximum error over the known samples \cite{muresan2005demosaicing}. 

In the remainder of this paper, first we review some aspects of causal inference. Then we  describe our approach in the context of panel data. The efficacy of our approach is lastly supported by artificial and empirical experiments. 

\section{Preliminaries}
To fix concepts and common terms let us give a brief overview of causal inference in the context of comparative case studies with panel data. Then we introduce the problem formulation and fix notations. Later, in Section \ref{sec:methods}, we describe our method.

Causal analysis takes a step further from standard statistical analysis, inferring beliefs under changing conditions to uncover causal relationships among variables \cite{pearl2009causalInferenc}. \textit{Causal inference} has been widely used in social and health sciences. 
Several frameworks to tackle causality analysis, such as structural models \cite{pearl2010causal} and the potential outcome framework \cite{rubin1974causaleffect} (which we focus on here) have been proposed. 

\subsection{Potential Outcomes Framework} This framework assumes effects are tied to a treatment or an intervention. To reveal the causal effects of an intervention, \cite{rubin1974causaleffect} proposed to measure the difference of two potential outcomes; let us denote them as $Y_N$ and $Y_I$, for a given unit $x$. The potential outcome $Y_N$ is the outcome for $x$ without being exposed to an intervention, and $Y_I$ is the outcome after an intervention is applied on $x$. So, the causal effect is 
\begin{equation}
    \tau = Y_N-Y_I \mbox{ .} 
\end{equation}
However, in real applications, we can never observe both outcomes for the same unit under the same conditions, only one of the two will take place at a given time. Therefore, one of the potential outcomes will always be \textit{missing}, and that is the core objective of the framework to estimate one of the outcomes. 

Let us introduce the main terms used in the potential outcomes literature, which are used throughout. A \textit{unit} is the atomic object in the framework, which can be any physical object, whether a patient, a city, or a collection of objects at a particular time. A \textit{treatment} is the action applied on a unit to change its state. The treatment\footnote{The terms \emph{treatment} and \emph{intervention} are used interchangeably.} can be a medicine given to a particular group or a lockdown order during a pandemic. The treatment is usually thought of as binary, so one group receives the treatment (the \textit{treated} group), and the other does not (the \textit{control} group).  Also, given an intervention at time $t_0$, there are two time periods: the \textit{pre-intervention/pre-treatment} period for $t < t_0$ and the \textit{post-intervention} period for $t > t_0$. 

Several works in econometrics and related literatures estimate the unknown potential outcome to estimate the causal effects. Here, we focus on the commonly used synthetic control method. 

\subsection{Synthetic Control Methods}
Synthetic control (SC) \cite{abadie2010synthetic} proposes a particular way to measure the missing observable potential outcome to estimate causal effects. In their evaluation of  econometrics for policy evaluation, \cite{atheyI2016state} asserted SC is ``arguably the most important innovation in the policy evaluation literature in the last 15 years''. Instead of using a single control unit or a simple average of a set of control units, SC creates a \textit{synthetic} unit to act as a control group by selecting appropriate weights for selected control units. The choice of weights should result in a synthetic control unit that best resembles the pre-intervention values of the treated unit. SC is based on a strong assumption that such weights exist if and only if the treated units' pre-treatment time series is inside the convex hull of the control units' time series. 

Therefore, SC is subject to the curse of dimensionality, in which the probability that exact matching of weights vanishes as the number of time periods grows \cite{zhao2017entropy}. In \cite{ferman2021synthetic}, the \textit{demeaned SC} (DSC) was proposed to relax SC constraints on weights to allow for a good pre-treament fit when the length of pre-intervention is very large. Also, SC tends to heavily depend on the selection of control units, and its estimates are biased by noisy control units.   \textit{Robust} SC \cite{Amjad18} (RSC) views the setting of SC as an instance of the Latent Variable Model. The observable outcomes of the control group are obtained by a low-rank approximation using singular value decomposition (SVD) and noise is eliminated. This also generalizes to cases where outcomes are missing.

\subsection{Problem Formulation}
Consider panel data\footnote{Panel data is another word for cross-sectional time-series data} having a collection of time series with respect to an aggregated metric of interest (e.g., country GDP). The data includes $N$ units observed over $T$ periods of time. Let $T_0$ be the intervention time, which splits the time period into a pre-intervention period with $ 1 \leq T_0 < T$ and a post-intervention period with length $T-T_0$. We fix $i=1$ ($i\in \{1,\dots, N\}$) for the treated unit at time $t$, hence, let $s_{1t} \in \mathbb{R}^{T}$ be the treated unit. The remaining units $i = 2, \ldots, N$ are the controls that are not affected by the intervention. Let $\mat{S} \in \mathbb{R}^{N-1\times T}$ be the control units matrix. 

Let the outcomes of the control and treated units follow a factor model, a common model in the econometrics literature \cite{abadie2021synthetic}.  
Let $x_{it}$ denote an aggregated metric for a unit $i$ at time $t$. In the absence of covariates and unobserved outcomes, the factor model is the following: 
\begin{equation}
    x_{it} = s_{it} + \epsilon_{it} \mbox{ .}
    \label{eq:factor_model}
\end{equation}
Following the literature in data imputation \cite{athey2021matrix,Amjad18} and optimal recovery for missing values \cite{muresaP2004}, we consider $s_{it} = \eta_i \psi_t $, where $\eta_i \in \mathbb{R}^{N-1}$ and $\psi_t \in \mathbb{R}^{T}$ are the latent features that capture the unit and time specifications respectively for observed outcomes, with independent random zero-mean noise, $\epsilon_{it}$. 
The goal is to approximate the partially-observed treated unit vector, $s_{1}$, to recover missing outcomes. 

To distinguish pre- and post-intervention periods, let $\mat{S} = [\mat{S}^-, \mat{S}^+]$, where $\mat{S}^-=\{s_{it}\}_{2\leq i \leq N, t\leq T_0}$, and $\mat{S}^+ = \{s_{it}\}_{2\leq i \leq N, T_0 < t \leq T}$. Vectors are defined in the same manner, i.e. $s_{i} = [s_i^+, s_i^-]$. 
The inverse of $\mat{A}$ is $\mat{A}^{-1}$. The Moore-Penrose pseudo-inverse of $\mat{A}$ is $\mat{A^{\dag}}$. The transpose of $\mat{A}$ is $\mat{A}^{\top}$. The $A$-norm of a vector $u$, denoted $||u||_A$, is the value of $u^{\top}\mat{A}u$. 

\section{Methods} 
\label{sec:methods}
In this section we describe our proposed method. In Section~\ref{sec:optimalrecovery}, we briefly introduce the \textit{optimal recovery} method from signal processing, which is a fundamental building block of our approach. In Section~\ref{sec:eopr} we describe our approach of estimation in the context of time-series panel data in comparative studies. 

\subsection{Optimal Recovery}
\label{sec:optimalrecovery}
Optimal recovery was introduced to effectively approximate a function known to belong to a certain signal class with limited information about it \cite{optimalestimation_book_1977}. It has been applied to estimate missing or corrupted pixels in images \cite{muresaP2004} and missing values in biological data \cite{chenV2019}. Optimal recovery estimates the missing value using a learned signal class and a set of known values, and  provides deterministic error bounds allowing the calculation of worst-case values at these bounds \cite{muresanP2001}. One way the signal class is constructed is using an ellipsoidal set of vectors that pass through a hyperplane. 
\begin{definition}[Ellipsoids] Given a matrix $Q$, an ellipsoid $K$ is a bounded convex set which has the form 
\begin{equation}
    K = \{x \in \mathbb{R}^n: x^TQx \leq h\}
    \label{eq:ellipsoid}
\end{equation}
where $Q=Q^T$, and $Q\succ0$. That is, $Q$ is symmetric and positive definite \cite{boyd2004convex}. 
\label{def:ellipsoid}
\end{definition}
The value $h$ is the radius of the ellipsoid. The matrix $Q$ determines how far the ellipsoid extends in every direction from the center. The lengths of the ellipsoid semi-axes are determined by the eigenvalues of $Q$. 

\subsection{Ellipsoidal Optimal Recovery (EOpR)}
\label{sec:eopr}
Here, we extend \textit{optimal recovery} with ellipsoidal signal class to estimate causal effects in panel data. Optimal recovery not only imputes the missing parts of the vector of interest but also provides deterministic worst-case characterization for the estimate. Under the potential outcomes framework, causal inference can be treated as a missing data problem, a kind of matrix completion \cite{athey2021matrix}. Therefore, we use optimal recovery from approximation theory to recover the missing outcomes for causal inference, in which optimal recovery has not been considered before.

\begin{figure}[!ht]
    \centering
    \includegraphics[width=0.70\linewidth]{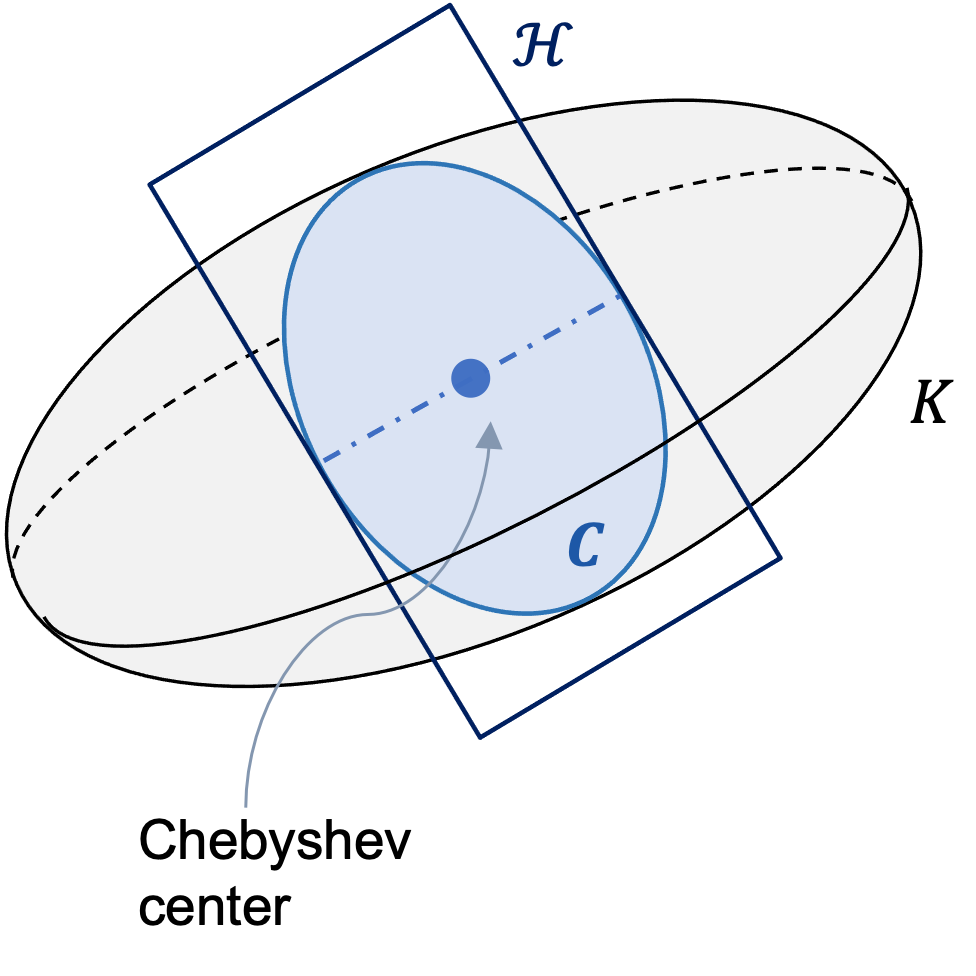}
    \caption{Geometric illustration for the optimal recovery algorithm with hyperplane $\mathcal{H}$ intersecting ellipsoid $K$, creating ellipse $C$, with a Chebyshev center}
    \label{fig:opr_illustration}
\end{figure}

Geometrically, we assume control units vectors in $\mat{S}$ belong to an ellipsoidal class $K$, and the pre-intervention of control units $\mat{S}^-$ belong to a hyperplane $\mathcal{H}$. We consider the treated unit $s_{1}$ as a partially-observed vector, where $s_{1}^+ = \{s_{1t}\}_{t>T_0}$ (in the post-intervention) is unknown and requires approximation. The vector $s_{1}$ lies in $C$, the intersection of $K$ and $\mathcal{H}$. The aim is to find an estimator that minimizes the worst-case error.
This is equivalent to finding the Chebyshev center of $C$ \cite{boyd2004convex}. Figure~\ref{fig:opr_illustration} illustrates the geometrical setting of the optimal recovery approach.

The Chebyshev center provides a minimax optimal solution for the recovery problem \cite{boyd2004convex}, shown in the following theorem. 
\begin{theorem}[Minimax Optimality] Let $C \in \mathbb{R}^n$ be an ellipse that represents an intersection of an ellipsoid and a hyperplane. Ellipse $C$ is a bounded and convex set with nonempty interior. A Chebyshev center is a point inside $C$ and is the minimal farthest distance from other points in $C$. For a given point $s_1 \in C$, the estimator $\hat{s}_1 \in C$ 
\begin{equation}
    \min_{\hat{s}_1} \max_{s_1 \in C} ||\hat{s}_1-s_1|| \mbox{.}
    \label{eq:cheby_center}
\end{equation}
\label{th:cheby_center}
is a Chebyshev center and it is the minimax estimator for $s_1$.
\end{theorem}
\begin{IEEEproof} Follows directly from the definition of the Chebyshev center \cite{boyd2004convex, optimalestimation_book_1977}.
\end{IEEEproof}

To find the minimax estimator $\hat{s}_1$, first construct the ellipsoidal class $K$ as in \eqref{eq:ellipsoid}, and then extrapolate using the representors of the known samples. 

\subsubsection{Learning the ellipsoid}
 We construct a covariance matrix $\mat{\Sigma}$ from the data matrix $\mat{S}$, such that $\mat{\Sigma} = \mat{S}\mat{S}^{\top} + \lambda \mat{I}$, where $\mat{\Sigma} \in \mathbb{R}^{T\times T}$, $\lambda$ is a scalar, and $\mat{I}$ is the identity matrix. 

To learn the ellipsoidal class $K$ in \eqref{eq:ellipsoid}, we want $K$ to have the most stretch in the same direction of $\mat{\Sigma}$, hence we let the eigenvalues of $\mat{Q}$ be the reciprocal of the eigenvalues of $\mat{\Sigma}$,
\begin{equation}
   \bm{Q = \Sigma^{\dag}} \mbox{ .}
   \label{eq:Q}
\end{equation}

\paragraph{Choice of parameter}
Based on the ellipsoid definition \eqref{def:ellipsoid}, the matrix $\mat{Q}$ must be positive definite. To ensure that eigenvalues of $\mat{Q}$ are strictly positive, we add a small perturbation $\lambda \in (0, 1]$ to the diagonal of $\mat{SS^{\top}}$, such that it minimizes the $\ell_2$-norm between outcomes in the pre-intervention period.

\paragraph{Learning the representors} 
 
From the pre-intervention vectors $\mat{\Sigma}^{-}$, from the covariance matrix $\mat{\Sigma}$, we derive the representors $\mat{\Phi} \in \mathbb{R}^{T_0 \times T_0}$, by the Riesz representation theorem \cite{muresaP2004} as
\begin{equation}
    \mat{\Phi} = \mat{\Sigma}^{-^{\top}}\mat{Q} \mat{\Sigma}^- \mbox{ .}
\end{equation}

\subsubsection{Extrapolation}
Now we calculate $\hat{s}_{1}$, the Chebyshev center of $C$. Given that the representor vectors lie in a subspace that is parallel to the center of the ellipse $C$,  $\hat{s}_1$ is a linear combination of the inverse of representors $\mat{\Phi}$, such that the optimal weights $w^{\ast}$ are
\begin{equation}
    w^{\ast} = (\mat{\Phi})^{-1}s_1^{-} \mbox{, }
    \label{eq:weights_extrapolation}
\end{equation}
then, the Chebyshev center, the estimated outcome $\hat{s}_1$ is
\begin{equation}
    \hat{s}_1 = \mat{\Sigma}^{\top} w^{\ast} \mbox{.}
    \label{eq:s_extrapolation}
\end{equation}

\subsection{Worst-case estimations}
Given that the estimator $\hat{s}_1$ is at the center of the intersection $C$, the vectors on the boundary of $C$ are the worst-case estimates. To attain worst-case vectors (minimax control of the estimates), let $y$ be the unit norm in $\mathcal{Z}$, a parallel subspace to representors $\mat{\Phi}$, determined by
\begin{equation}
    y = \mat{\Phi}^{-1} \mat{\Sigma}^{-^{\top}} \mat{Q} \mat{\Sigma}\mbox{.}
\end{equation}
The worst case estimates $\bar{s}_1$ are: 
\begin{equation}
    \bar{s}_1 = \hat{s}_1 \pm (\varepsilon - \|\hat{s}_1\|_Q)^{\frac{1}{2}}y,
\end{equation}
with a very small $\varepsilon$, and the $\mat{Q}$-norm $\|\hat{s}_1\|_Q = \hat{s}_1^{\top} Q \hat{s}_1$.

\subsection{Properties of the estimator}
Based on the optimal recovery approach, the extrapolation step produces an estimator that is the Chebyshev center of the ellipse $C$ with desirable properties of unbiasedness and consistency.
\begin{theorem}
    Let $r$ denote the rank of matrix $\mat{\Sigma}$, for $\lambda \geq 0$, the estimation error can be bounded as 
    \begin{equation}
        \mathit{MSE}(s_1, \hat{s_1}) \leq \frac{2\sigma^2 |r|}{T}
    \end{equation}
such that as $T\to \infty$ the algorithm produces a consistent estimator.  
\end{theorem}
\begin{IEEEproof} In Appendix~\ref{apx:analytical_proof}. 
\end{IEEEproof}

The Chebyshev center has been also proved to be a consistent and unbiased estimator geometrically \cite{halteman1986chebyshev}, full proof in Appendix~\ref{apx:geometric_proof}. 
\section{Experiments}
We compare the accuracy of our ellipsoidal optimal recovery (EOpR) approach against other causal inference methods used in policy evaluation: SC \cite{abadie2003economic}, RSC \cite{Amjad18}, DSC \cite{ferman2021synthetic}, and SDID \cite{Arkhangelsky21}. We first evaluate on simulated data to demonstrate properties of EOpR under certain settings. We then  evaluate on two classical panel datasets commonly used in the SC literature (California Proposition 99 \cite{abadie2010synthetic} and Basque Country \cite{abadie2003economic}). We finally apply EOpR in the context of the COVID-19 pandemic to estimate the number of confirmed cases in New York State. 

\subsection{Evaluation Metrics} To measure the quality of estimation, we use two metrics. First, we measure the root-mean-square error (RMSE) of estimated signals. The pre-intervention (training) error is for $1 \leq t \leq T_0$, and a post-intervention (testing) error is for $T_0 < t \leq T_0$  
\begin{equation}
    \text{RMSE}(u, \hat{u}) = \left(\frac{1}{\mathcal{T}}\sum_{t=1}^{\mathcal{T}}(u - \hat{u})^2\right)^{1/2},
\end{equation}
where $\mathcal{T}$ is the size of the selected time period. 

Second, Abadie \cite{abadie2010synthetic} proposed a test statistic to evaluate the reliability of the estimates by running \textit{placebo tests}. One placebo test considers one control unit as a placebo treated unit and apply the estimation algorithm. Since control units are assumed to not be affected by the examined intervention, one would expect that the estimated signal for the placebo unit does not diverge from its corresponding control unit. Further, the gaps between each placebo estimation and its corresponding control unit should be less divergent than the gap between the original treated unit and its estimation. Placebo tests are applied on the classical econometrics case studies.

\subsection{Simulations}
We conduct artificial simulations to demonstrate the properties of EOpR estimates in both the pre- and post-intervention periods. We show that EOpR performs well and better than existing causal inference methods under various settings. 
\paragraph{Experimental setup} Consider a data generating process, which is frequently considered for low-rank matrix decomposition solutions, similar to \cite{Amjad2019_mRSC}, as follows. First we create two sets of row and column features, $B_r$, $B_c$, where $B_r = \{b_k|b_k \sim \text{Unif}(0,1), 1 \leq k \leq 10\}$ and $B_c = \{b_k|b_k \sim \text{Unif}(0,1), 1 \leq k \leq 10\}$. For each unit $2 \leq i \leq N$, we assign a parameter $\theta_i$ drawn from $B_r$ (with replacement), and for each time $1 \leq t \leq T$ we assign a parameter $\rho_t$ drawn from $B_c$ (with replacement). We use the following formula to generate a data point $\tilde{s}_{it} = f(\theta_i, \rho_t)$ to construct the control units
\begin{equation}
    f(\theta_i, \rho_t) = \frac{10}{1+\exp{(-\theta_i - \rho_t - (\theta_i\rho_t))}} + \epsilon_{it}, 
\end{equation}
where $\epsilon_{it} \sim \mathcal{N}(0,1)$, an independent Gaussian noise.
To construct the treated unit $\tilde{s}_{1t}$ of interest, we generate the vector using a uniform linear combination of row vectors. 

In the following experiments, we investigate EOpR resistance to bias in comparison to other algorithms under different sizes of units $N$ and time periods $T_0$, $T$. For each combination of $N$, $T_0$, and $T$ we generate $10$ simulations and average the resulting RMSE scores for the estimated pre- and post-intervention signals of $\tilde{s}_{1t}$.

\subsubsection{Length of pre-intervention period} Consider when the number of pre-treatment units vary proportionally to $T$, fix the size of units $N$ and the total size of time $T$. A short period of $T_0$ has shown to fail in reproducing the trajectory of the treated unit \cite{abadie2021synthetic}. Therefore, we vary the time of intervention $T_0$ between 10\% to 90\% of the entire time $T$. 

\begin{figure}[!ht]
    \centering
    \subfloat[Pre-intervention Error]{
\includegraphics[width=0.75\linewidth]{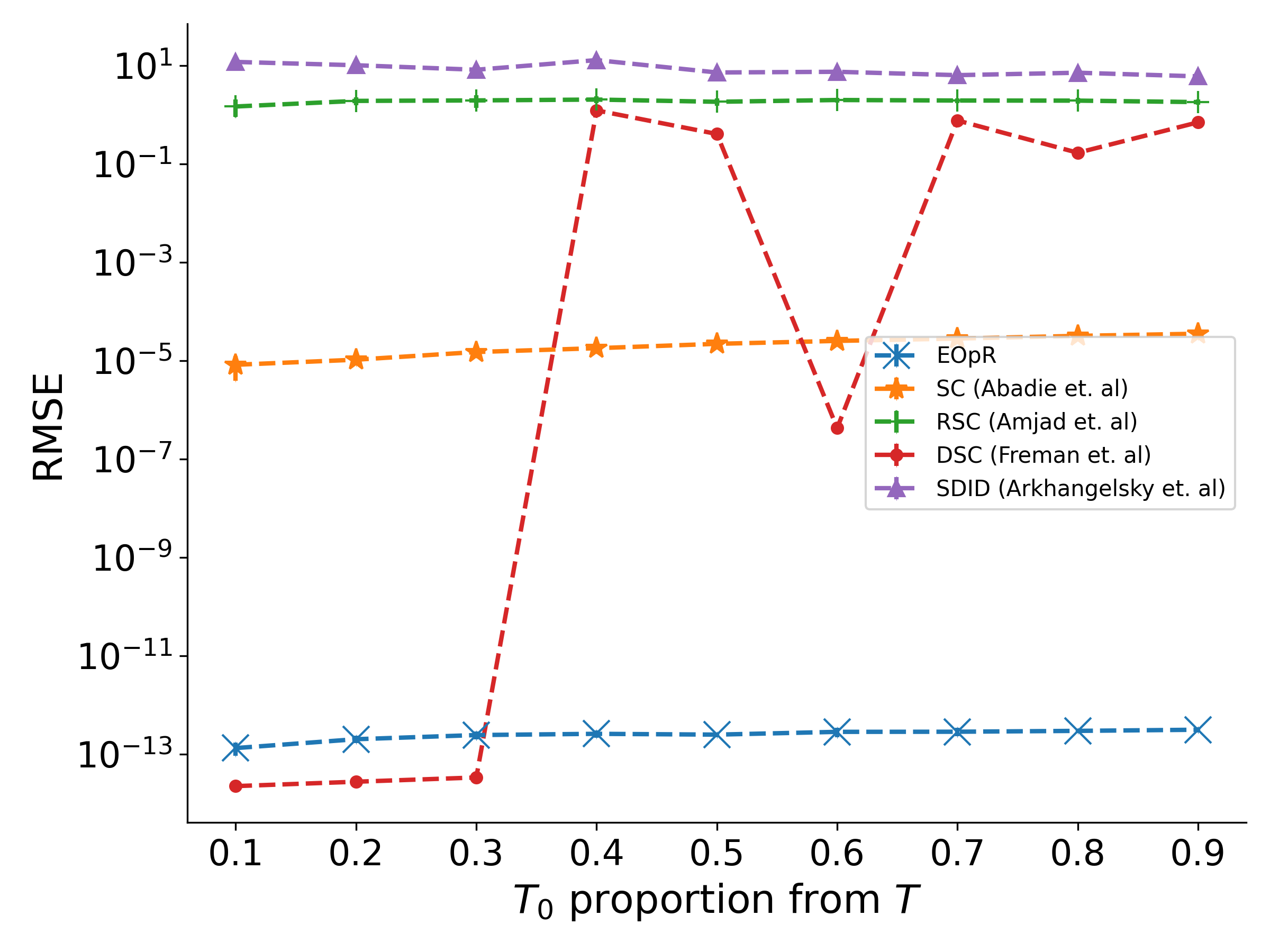}}
\vfill
    \subfloat[Post-intervention Error]{
    \includegraphics[width=0.75\linewidth]{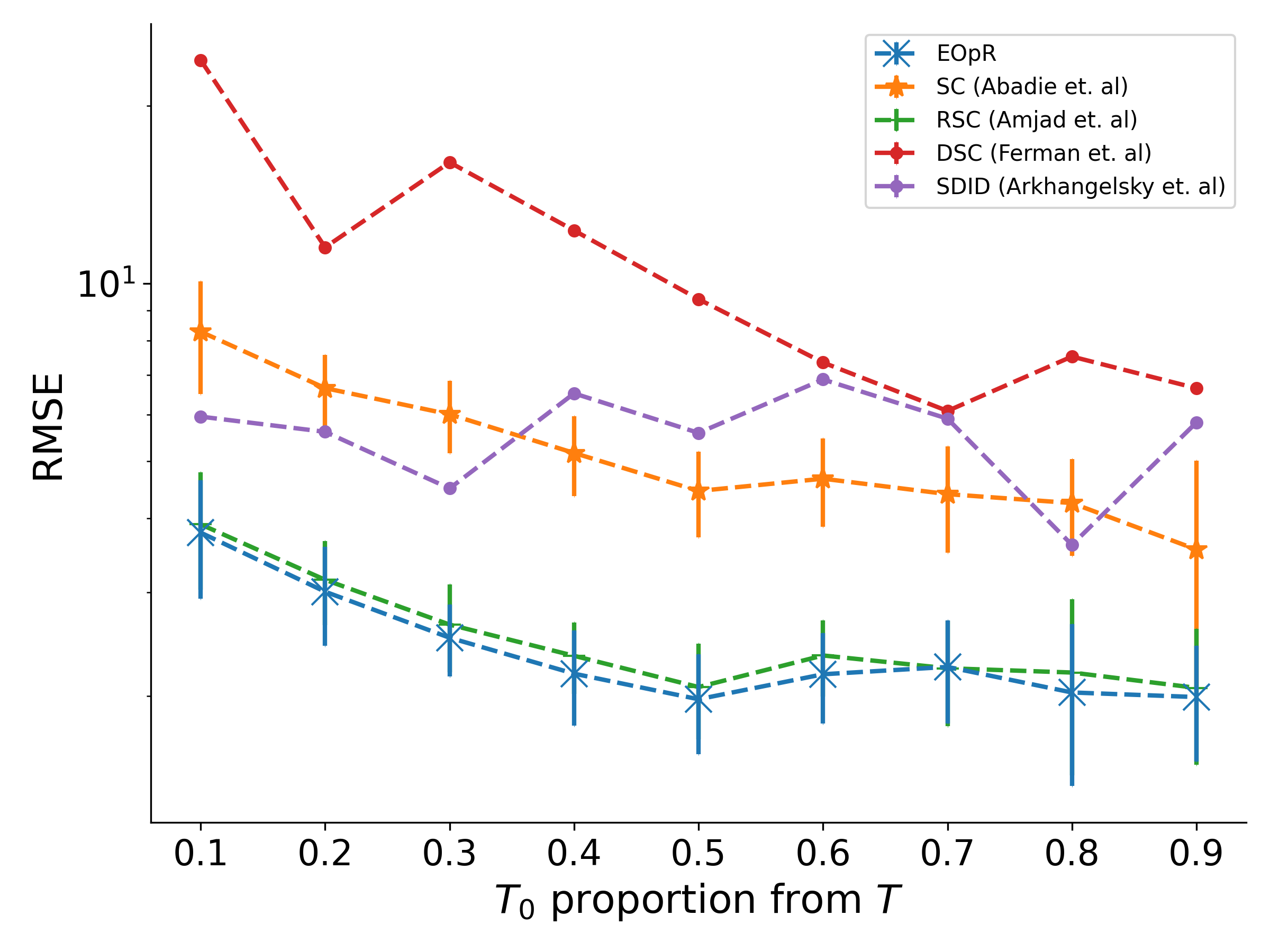}}
    \caption{Percentage of $T_0$ length of the total $T$, with $N=100$, and total $T = 50$}
    \label{fig:simulation_T0_increase}
\end{figure}

Figure~\ref{fig:simulation_T0_increase} shows the effect of different pre-treatment lengths on the algorithm's ability to estimate, with $N=50$, and $T=200$. When having either a small number or a large number of pre-treatment periods (relative to $T$), EOpR recovers the original treated signal with the minimum error compared to other algorithms. Specifically at 10\% and 20\% of $T$, where the pre-treatment period is short, EOpR  extrapolates beyond the training periods with the lowest bias in the estimation of the post-intervention trend, whereas other algorithms have higher bias in the post-intervention estimation. 

Additionally, at lengths of 40\% and onwards, the number of pre-treament vectors are greater than the size of $N$, $(N \ll T_0)$, a common setting adapted in \cite{abadie2010synthetic}, EOpR still maintains a lower bias in the post-intervention with consistent low training error. 

\subsubsection{Number of control units} With a large number of units, the risk of overfitting increases, which produces a high potential of increased estimation bias \cite{abadie2010synthetic}. Here, we model the increase of the number of units with fixed $T_0 = 25$ and total $T = 125$. Large number of control units, $N$, has shown to be challenging for SC as it exacerbates the bias of the estimator \cite{abadie2021synthetic}. 

\begin{figure}[!ht]
    \centering
    \subfloat[Pre-intervention Error]{
\includegraphics[width=0.75\linewidth]{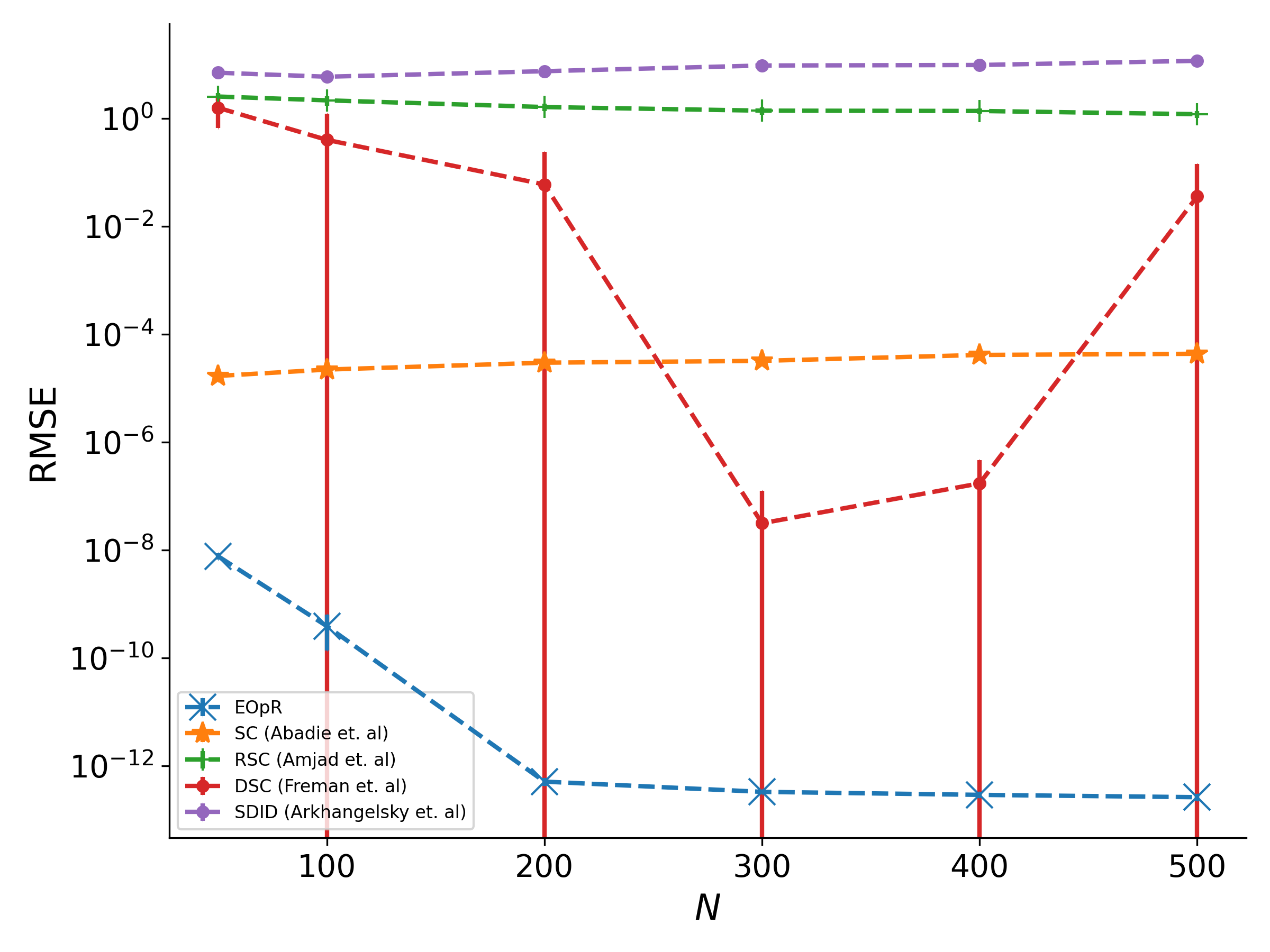}}
\vfill
    \subfloat[Post-intervention Error]{
    \includegraphics[width=0.75\linewidth]{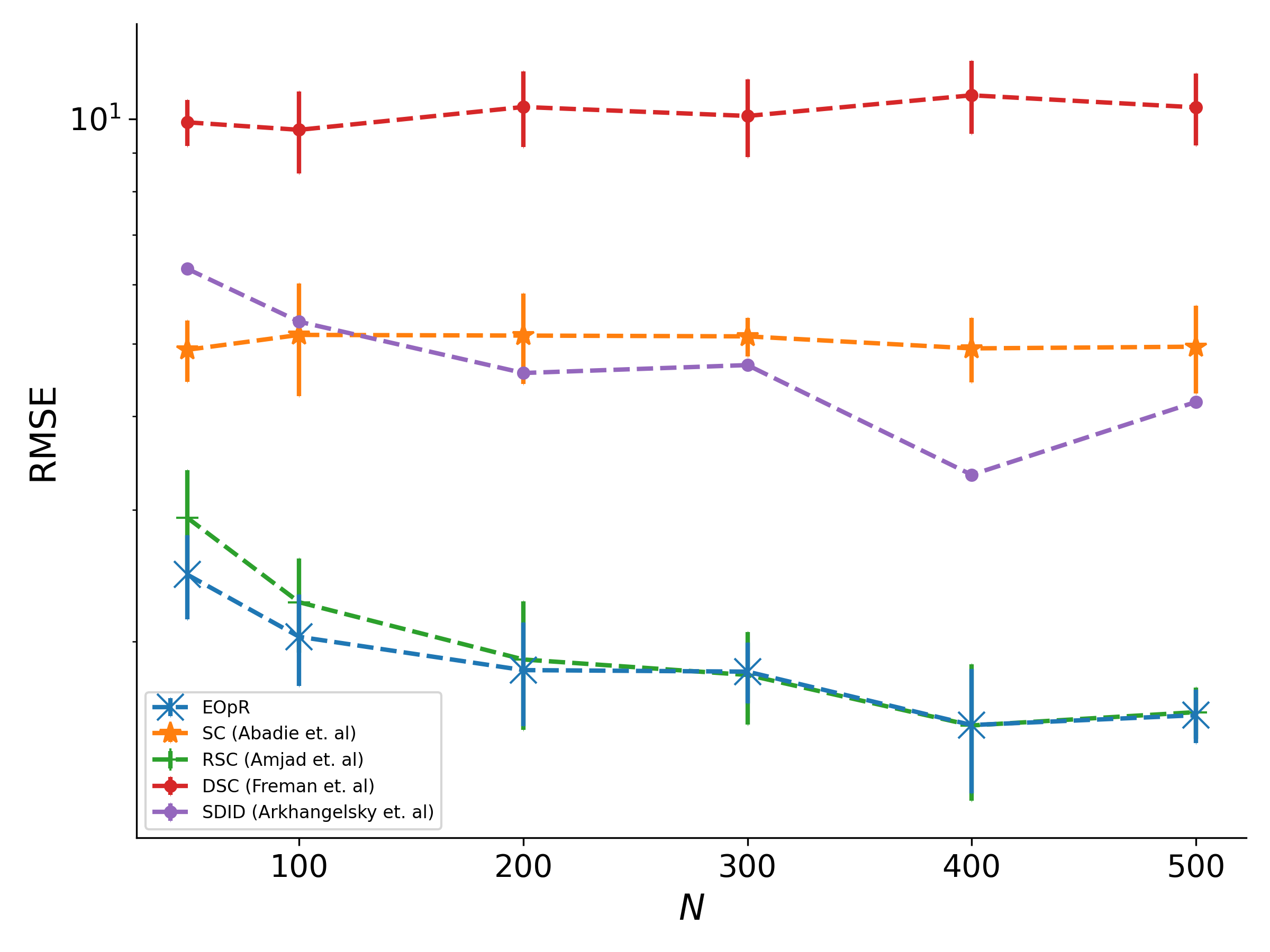}}
    \caption{Growing size of $N$, with $T_0=25$, and post-intervention $T = 100$}
    \label{fig:simulation_N_increase}
\end{figure}

Figure ~\ref{fig:simulation_N_increase} shows the effect of a growing number of units $N$ on  algorithm performance. Given that EOpR also achieves low error at the post-intervention estimation at greater sizes of $N$, EOpR has the ability to reconstruct the true signal trend even with large control units and potentially noisy settings.

\subsubsection{Length of post-intervention period} We further investigate the ability of EOpR to estimate the trajectory of the post-intervention for an extended period of time. The ability to estimate for an extended period indicates an algorithm is robust against estimation bias. 

\begin{figure}[!ht]
    \centering
    \subfloat[Pre-intervention Error]{
\includegraphics[width=0.75\linewidth]{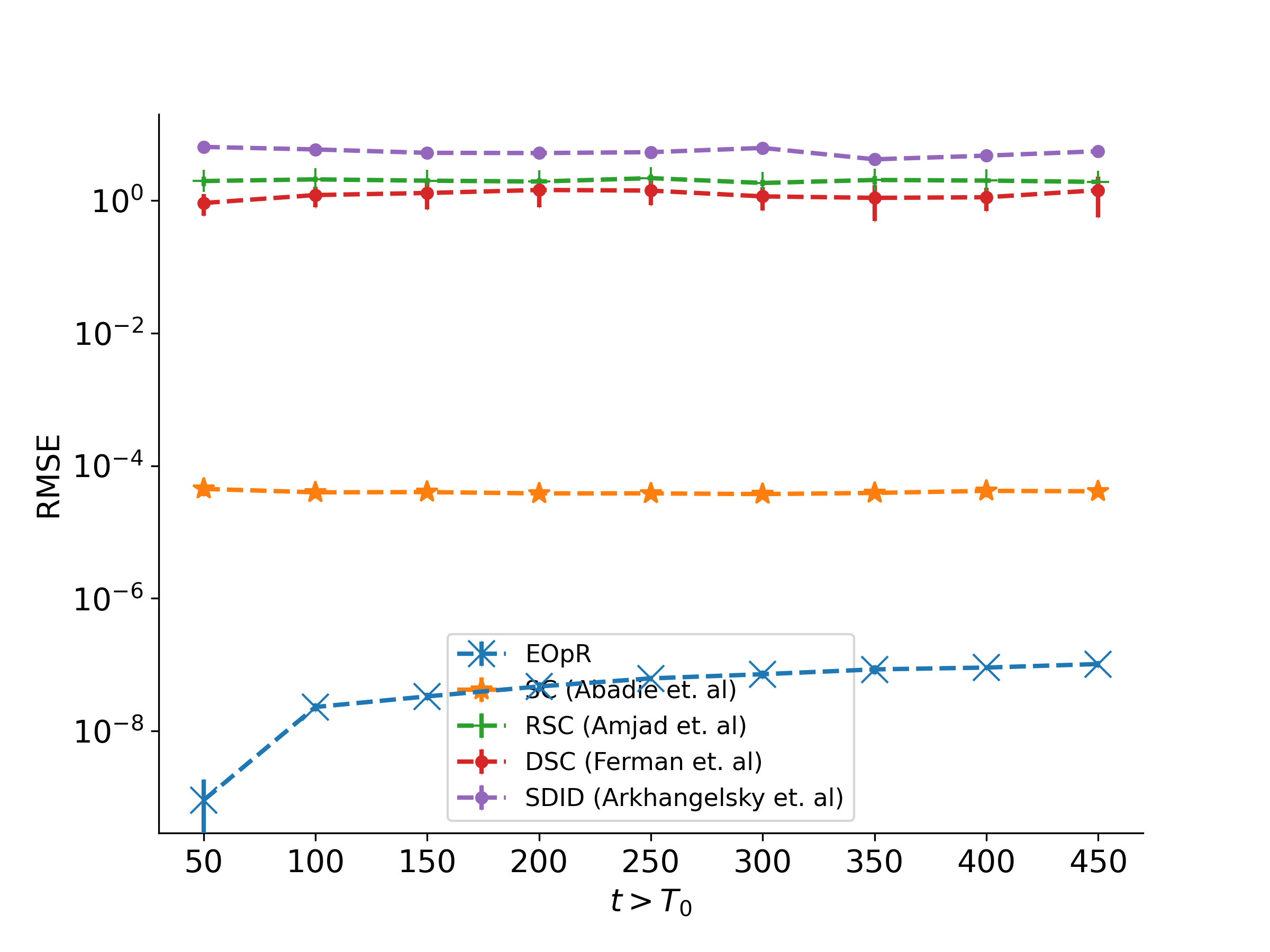}}
\vfill
    \subfloat[Post-intervention Error]{
    \includegraphics[width=0.75\linewidth]{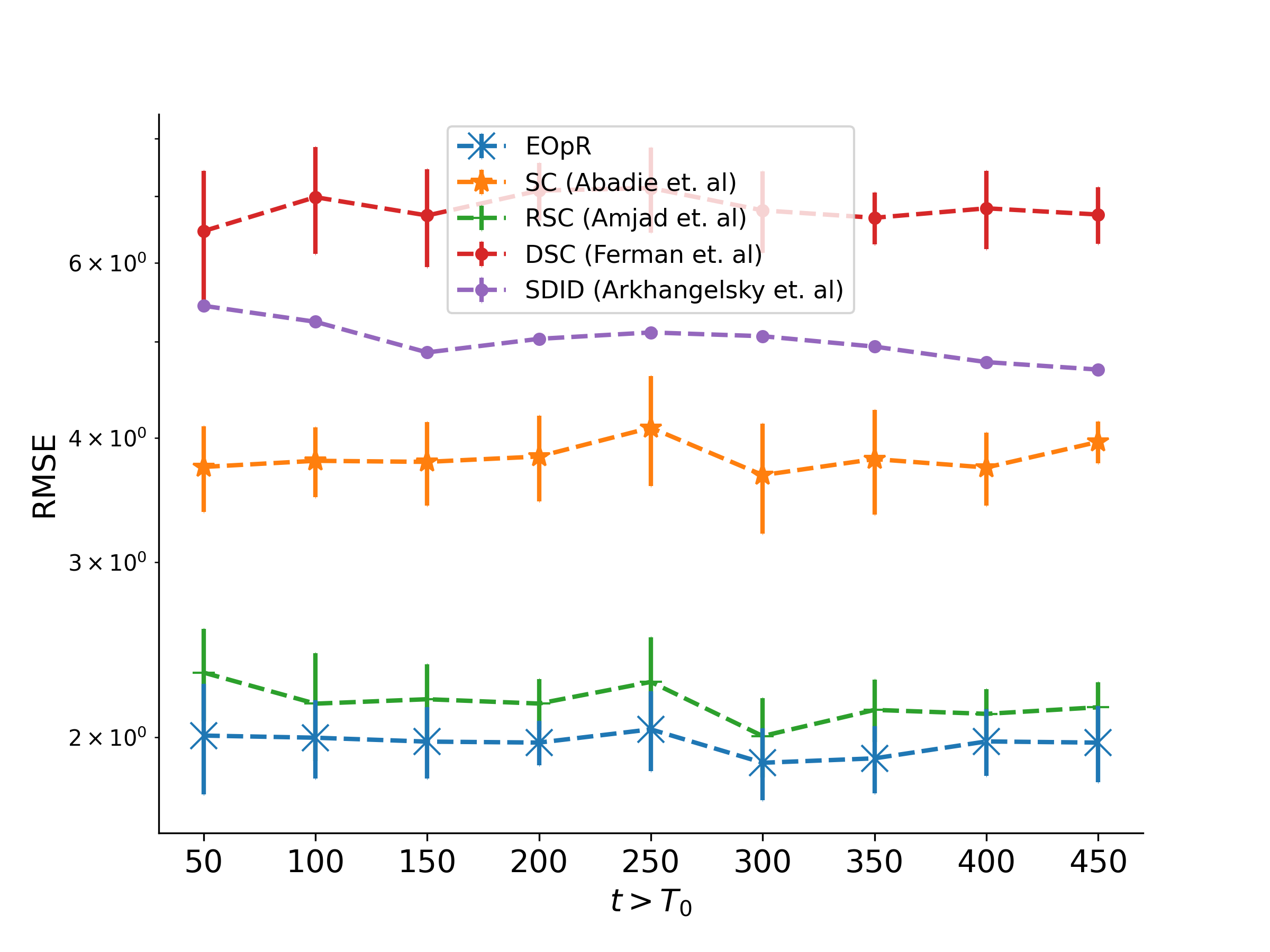}}
    \caption{Growing post-intervention periods, with $N=100$ and $T_0 = 50$}
    \label{fig:simulation_T1_increase}
\end{figure}

Figure~\ref{fig:simulation_T1_increase} shows the estimation errors with fixed $N=100$ and $T_0 = 50$, and tested over multiple lengths of post-intervention. EOpR consistently achieves a low estimation error in both pre- and post-intervention estimation, especially at longer periods of time, e.g.\ $(t=450)$.

\subsection{Real-World Experiments}
\subsubsection{Basque Country}
\begin{figure}[!ht]
    \centering
    \subfloat[Comparison of methods]{
\includegraphics[width=0.75\linewidth]{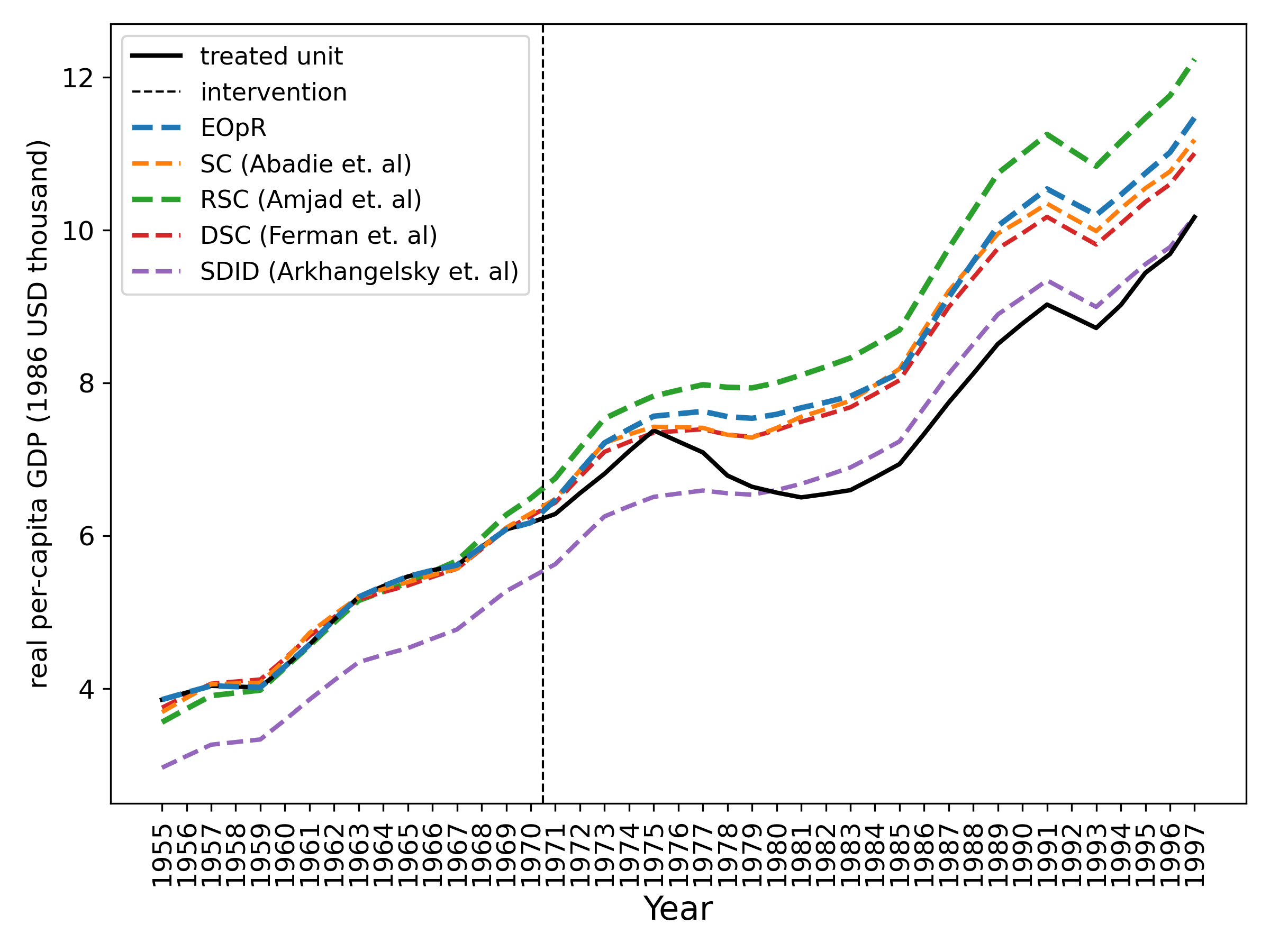}}
\vfill
    \subfloat[Solution with worst cases]{
    \includegraphics[width=0.75\linewidth]{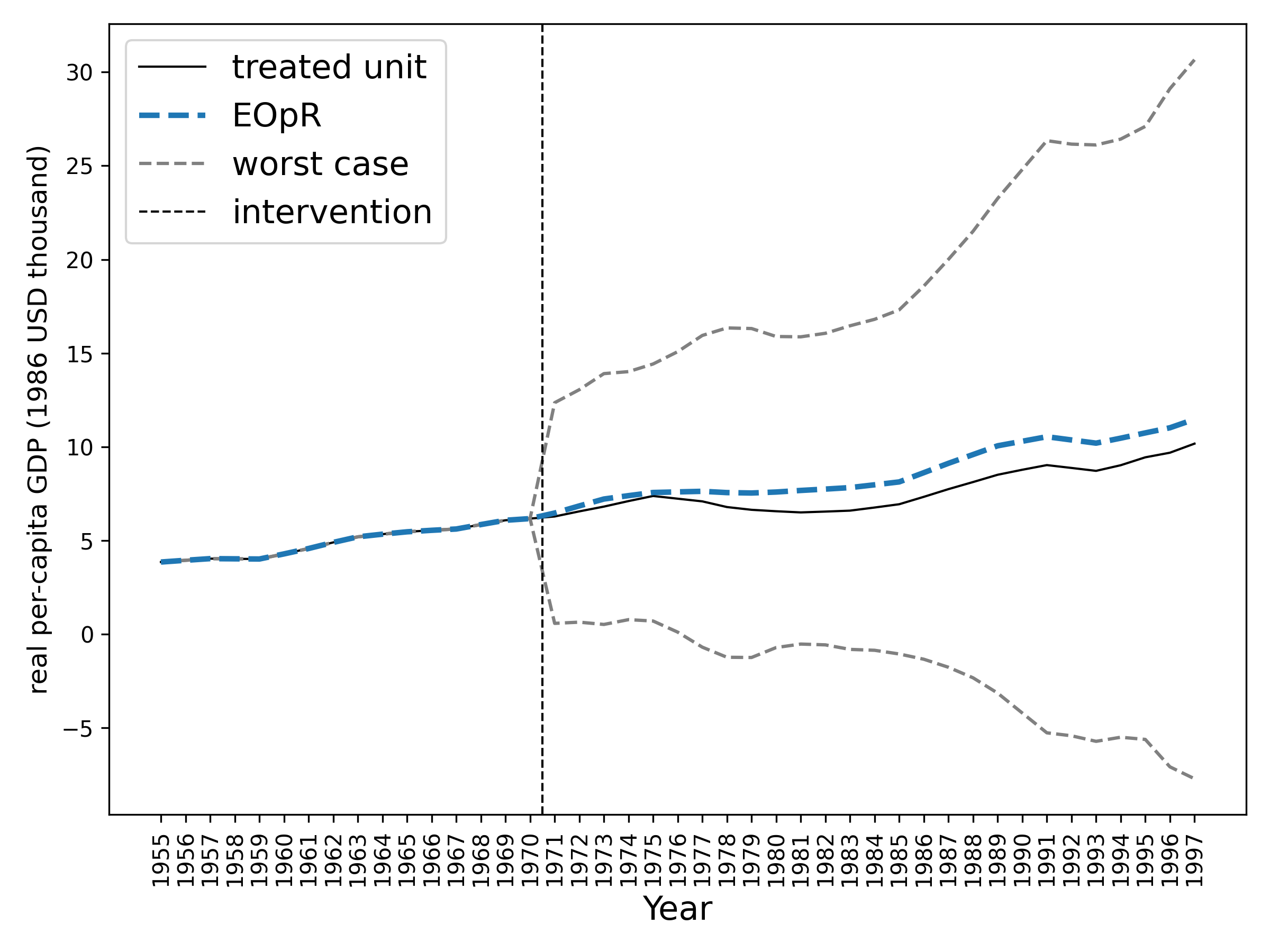}}
    \caption{Trends in per-capita GDP between Basque Country vs. synthetic Basque Country}
    \label{fig:basque_exp}
\end{figure}

The objective of this case study is to investigate the effect of terrorist attacks on the Basque Country economy compared to other Spanish regions. Terrorist activities started by 1970. There was a significant negative impact on the economy of Basque Country measured by per-capita GDP. Abadie \cite{abadie2010synthetic} showed that the economy would be better without terrorism.  
\paragraph{Results} Figure \ref{fig:basque_exp} shows the actual trajectory of the Basque Country economy in black, with a degradation after 1970. In comparison to other methods, our EOpR method recovers Basque Country estimate, with more accurate fit on the pre-intervention values of the treated unit. Figure \ref{fig:basque_exp} shows the estimated worst-case potential outcomes from EOpR. 

\begin{figure}[!ht]
    \centering
    \includegraphics[width=0.75\linewidth]{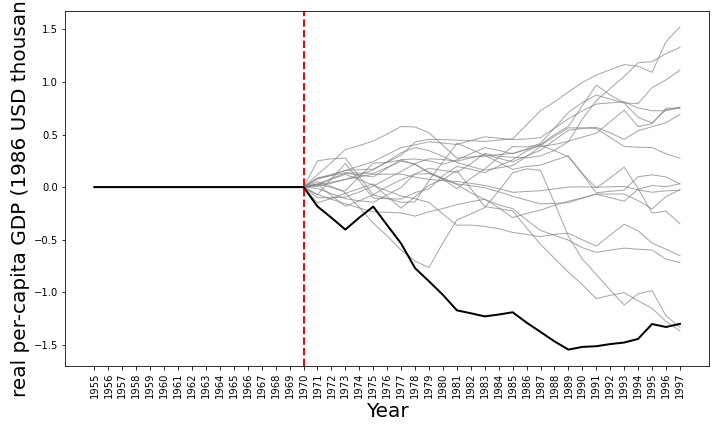}
    \caption{Placebo tests including all regions}
    \label{fig:basque_placebo}
\end{figure}

\paragraph{Placebo Tests} We create placebo tests, similar to Abadie \cite{abadie2010synthetic}. Note that in \cite{abadie2010synthetic}, authors excluded 5 regions which had poor fit in the pre-intervention, but we keep all regions. We plot the differences between our estimates and the observations of all regions as placebo and Basque Country (the actual treated unit). Figure \ref{fig:basque_placebo} shows the differences for all regions compared to Basque Country (solid black line in the figure). The divergence for Basque Country was the largest, thus, the derived estimates by the EOpR are reliable. 

\subsubsection{California Proposition 99}
The objective of this case study is to investigate the anti-tobacco legislation, Proposition 99, on the per-capita cigarette consumption in California in comparison to other states in the United States. The legislation took place in 1970. Without such legislation, the consumption of cigarettes in California would not have  decreased \cite{abadie2010synthetic}. 

\begin{figure}[!ht]
    \centering
    \subfloat[Comparison of methods]{
\includegraphics[width=0.75\linewidth]{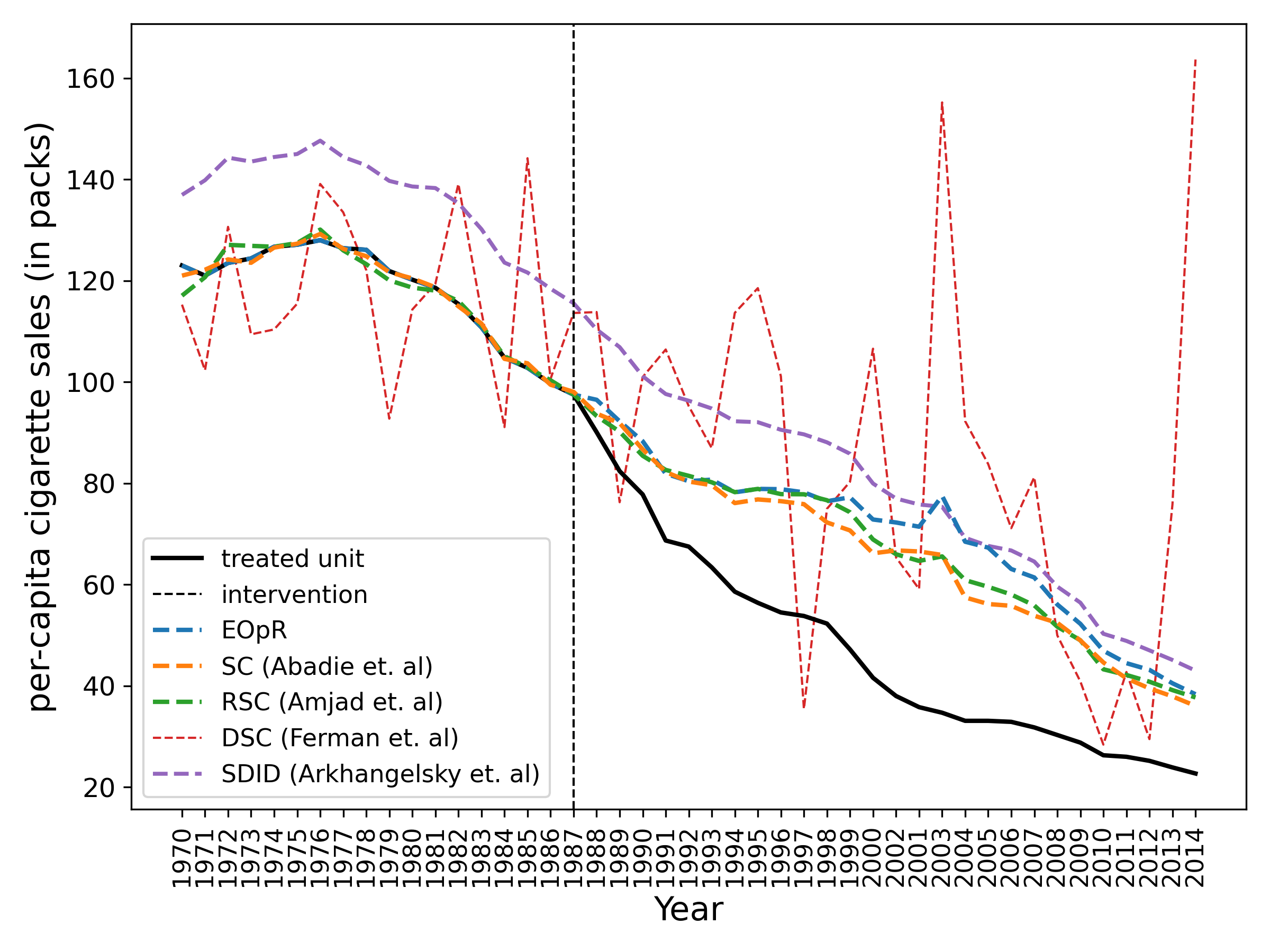}}
\vfill
    \subfloat[Solution with worst cases]{
    \includegraphics[width=0.75\linewidth]{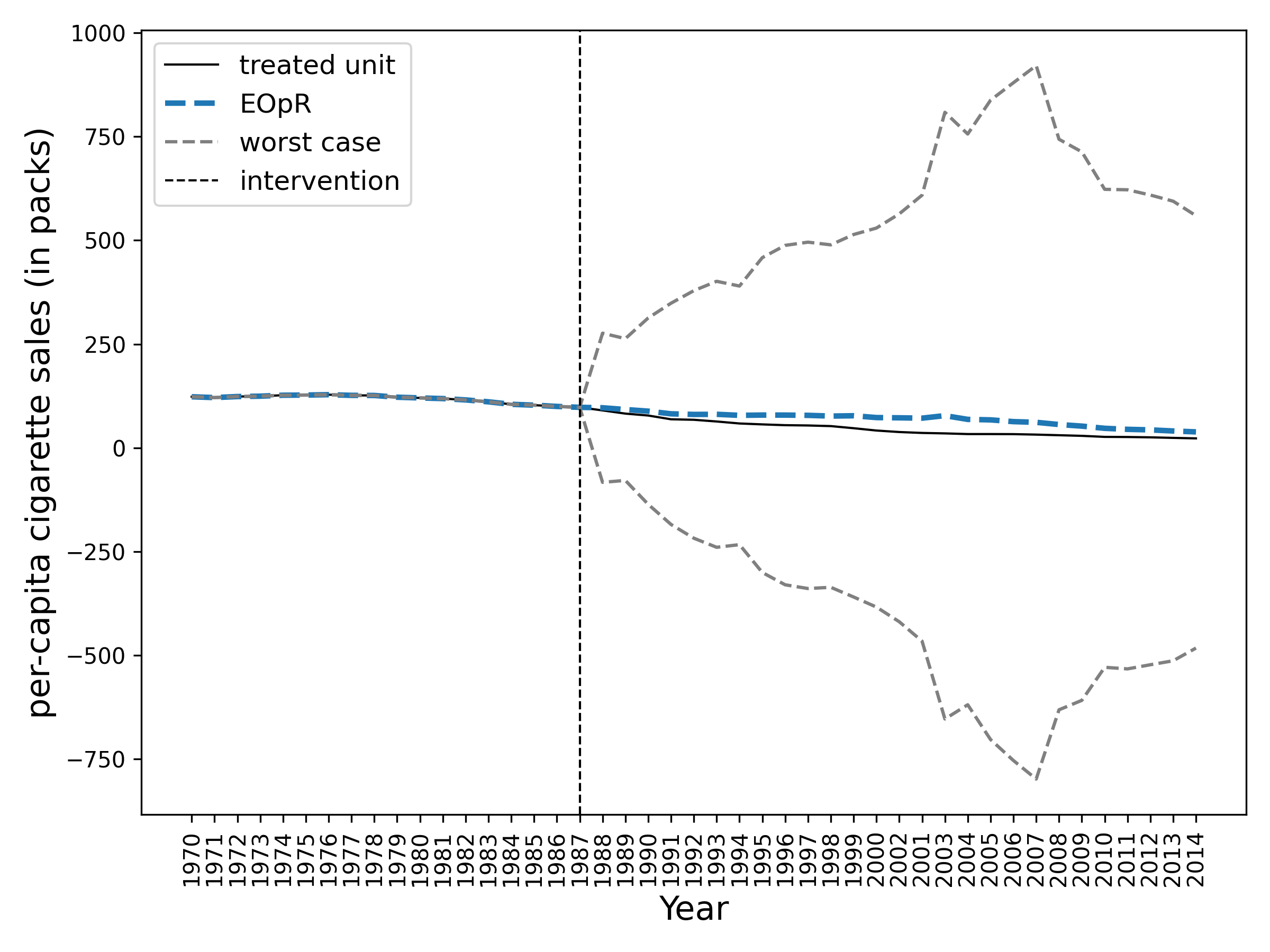}}
    \caption{{Trends in per-capita cigarette sales between California vs. synthetic California}}
    \label{fig:cali1}
\end{figure}

\begin{figure}[!ht]
    \centering
    \includegraphics[width=0.75\linewidth]{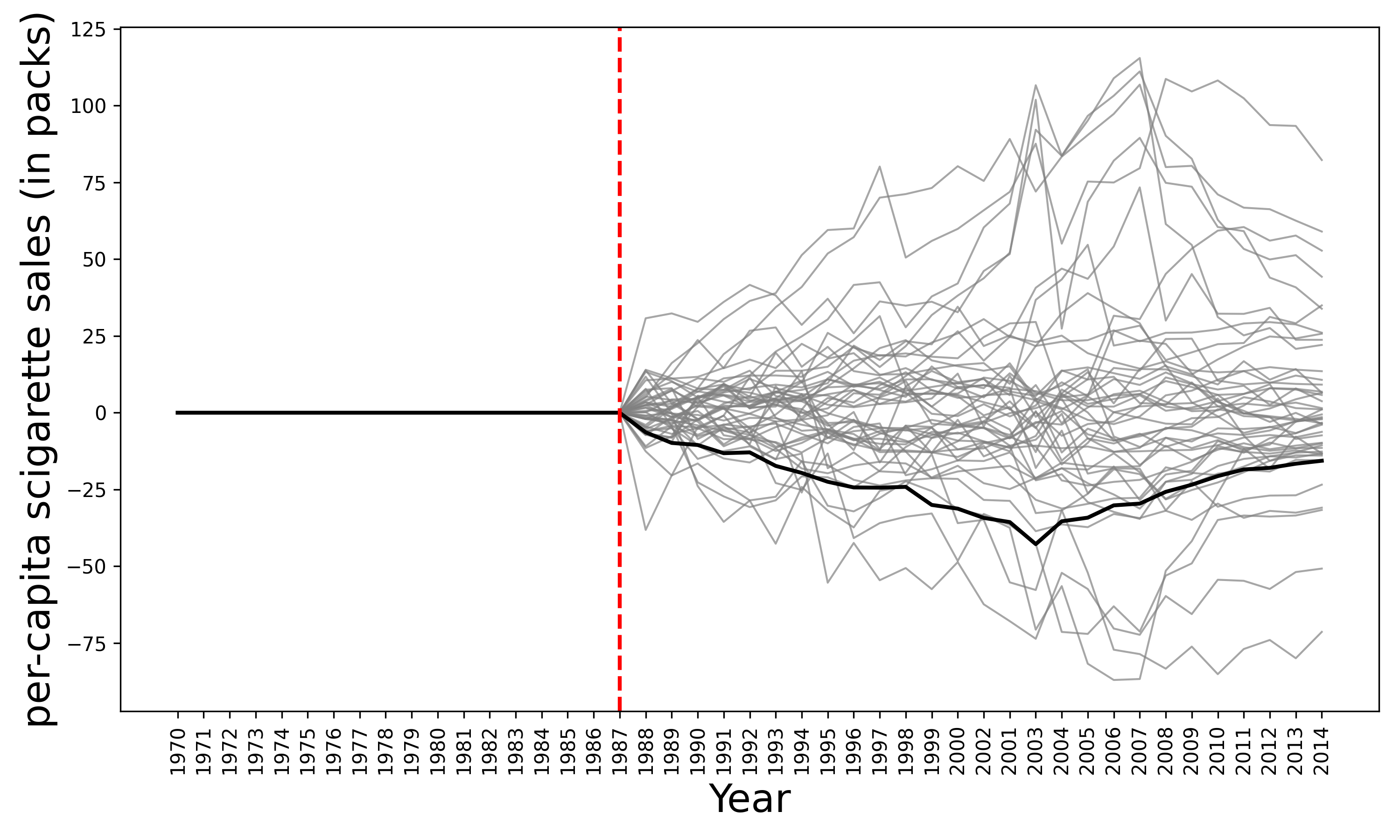}
    \caption{Placebo tests including all regions}
    \label{fig:cali_placebo}
\end{figure}

\paragraph{Results}
Figure \ref{fig:cali1} shows the actual trajectory of California cigarette consumption in black. Our method recovers the estimated signal better than other methods with an adequate fit on the pre-intervention outcome, and also it derives the worst-case estimates.  
\paragraph{Placebo Tests} We  apply the same placebo test to the California case study. Abadie \cite{abadie2010synthetic} excluded 12 regions, but we keep all of them. Figure \ref{fig:cali_placebo} shows the divergence between estimations and observations of all regions with solid black line for California. This shows a similar observation as in \cite{abadie2010synthetic}. 

=

\subsubsection{COVID-19 in New York}
New York was one of the earliest American states that turned to an epicenter of COVID-19 during 2020 \cite{thompson2020covid}. We aim to estimate the New York COVID-19 cases trajectory with states as control units using EOpR and other methods. To make the states uniform, we select the states where lockdowns were imposed, and end up with a total of 43 states. 
\paragraph{Experimental setup} For lockdown dates, we use data from the COVIDVis project \footnote{\url{https://covidvis.berkeley.edu/##lockdown_section}} that tracks policy interventions at the state level. We consider the dates of \textit{shelter-in-place} mandate. For COVID-19 cases, we consider state-level case load data from the New York Times database \cite{nytiems_cov}. Note that since reported cases depend on testing, our analysis is limited by the fact there was widespread shortage of available tests in different regions at different times. 

\begin{figure}[!ht]
    \centering
    \subfloat[States with relatively similar trends\label{fig:similar_ny}]{
\includegraphics[width=0.75\linewidth]{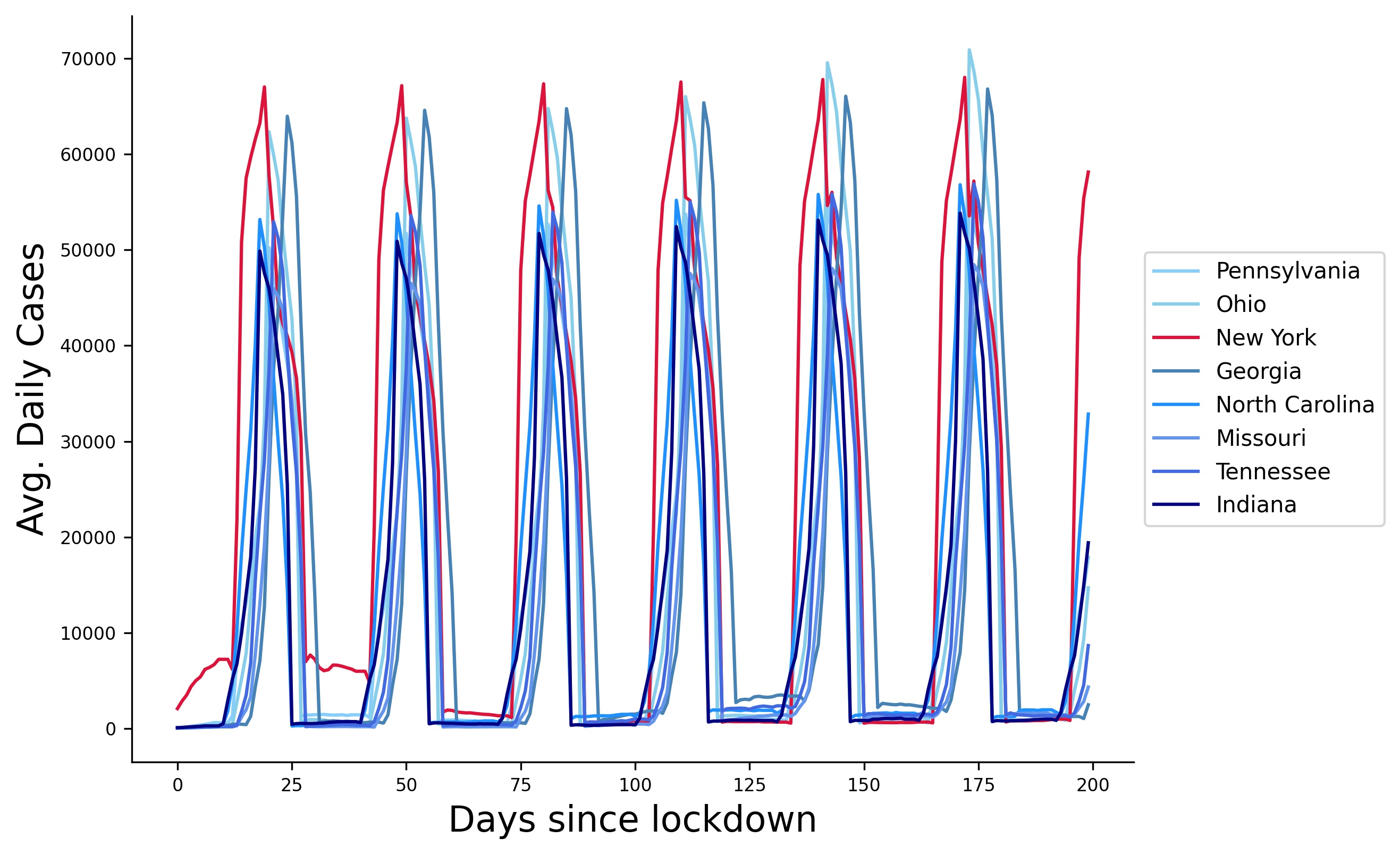}}
\vfill
    \subfloat[States with dissimilar trends\label{fig:dissimilar_ny}]{
    \includegraphics[width=0.75\linewidth]{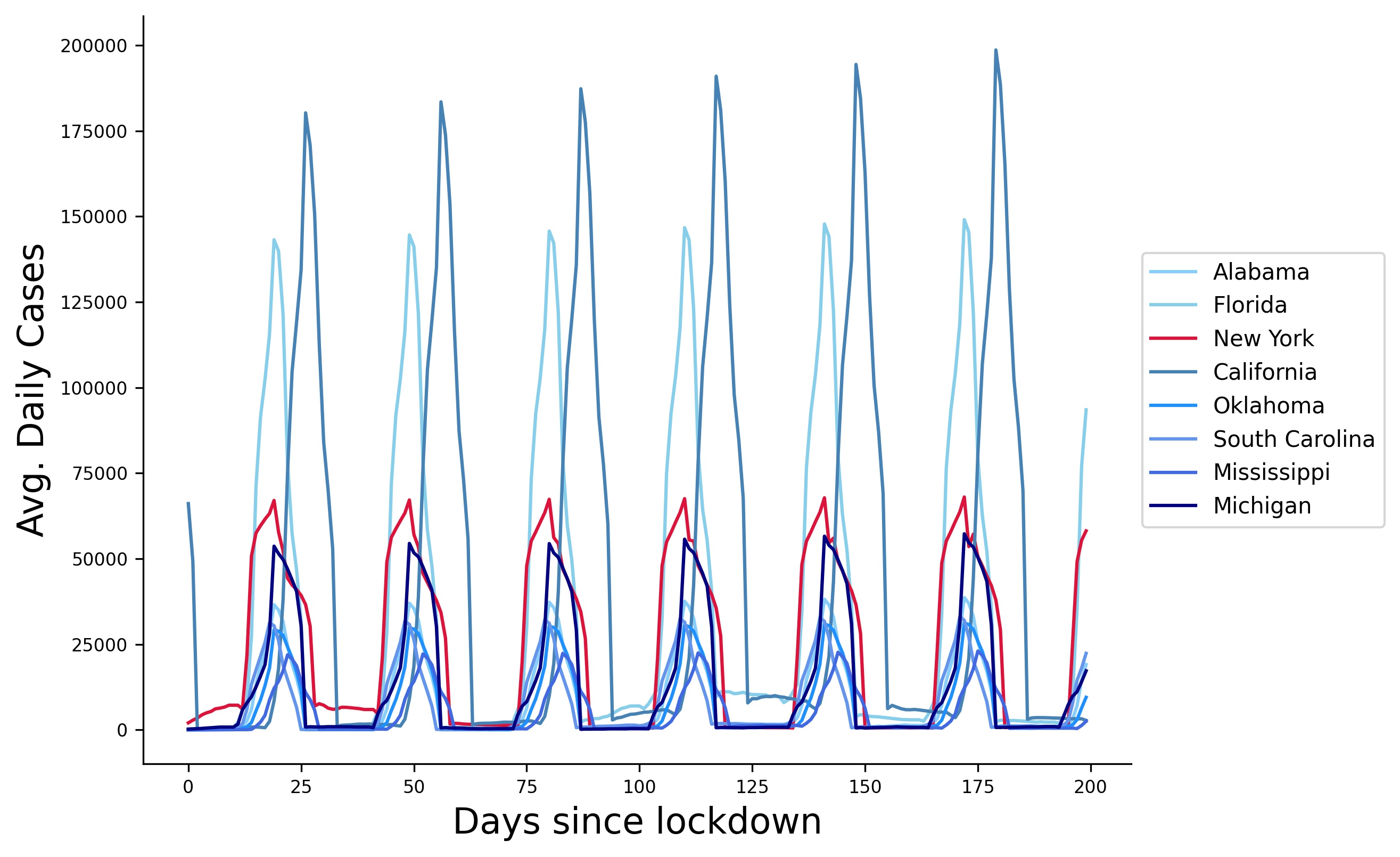}}
    \caption{{Moving average of COVID-19 confirmed cases of New York (in red) and 7 other states}}
    \label{fig:ny_trends}
\end{figure}

Since each state imposed a lockdown at different times, we aligned states based on the days differences between the time a lockdown took place and the rest of the dates during the period of interest, following \cite{bayat2020synthetic}. Figure \ref{fig:ny_trends} shows the trend of New York and 7 other states and their moving averages (over 7 days).

Based on the literature of SC \cite{abadie2010synthetic}, one would need to select the control units based on their similar trends to the target unit to ensure a correct construction of the treated unit. In the following experiments, we estimate NY COVID-19 cases using $\{5, 7, 15, 43\}$ states, for a period of 200 days, $T=200$, which approximates the period between March and August 2020, and time of intervention $T_0 = 50$. 
\paragraph{Results}
Table~\ref{tab:covid_movingAVG_errors} shows the errors produced by the four algorithms for pre- and post-intervention estimation. We start by $N=5$, and pick the states that have a very similar trend to the daily average of New York State, similarly for $N=7$ shown at Figure \ref{fig:similar_ny}. For a large $N$, EOpR recovers the trajectory of New York with the lowest estimation error regardless of the existence of noisy control units which are highly dissimilar to the New York COVID-19 trend. An example of some dissimilar states used is shown in Figure \ref{fig:dissimilar_ny}. The ability of EOpR to recover the true trajectory under different sizes and similarity of control units shows its potential to learn a set of high-quality signals that are sufficient for recovery.
\input{tables/covid_movingAVG}

\begin{figure}[!ht]
    \centering
    \subfloat[$N=5$]{
\includegraphics[width=0.7\linewidth]{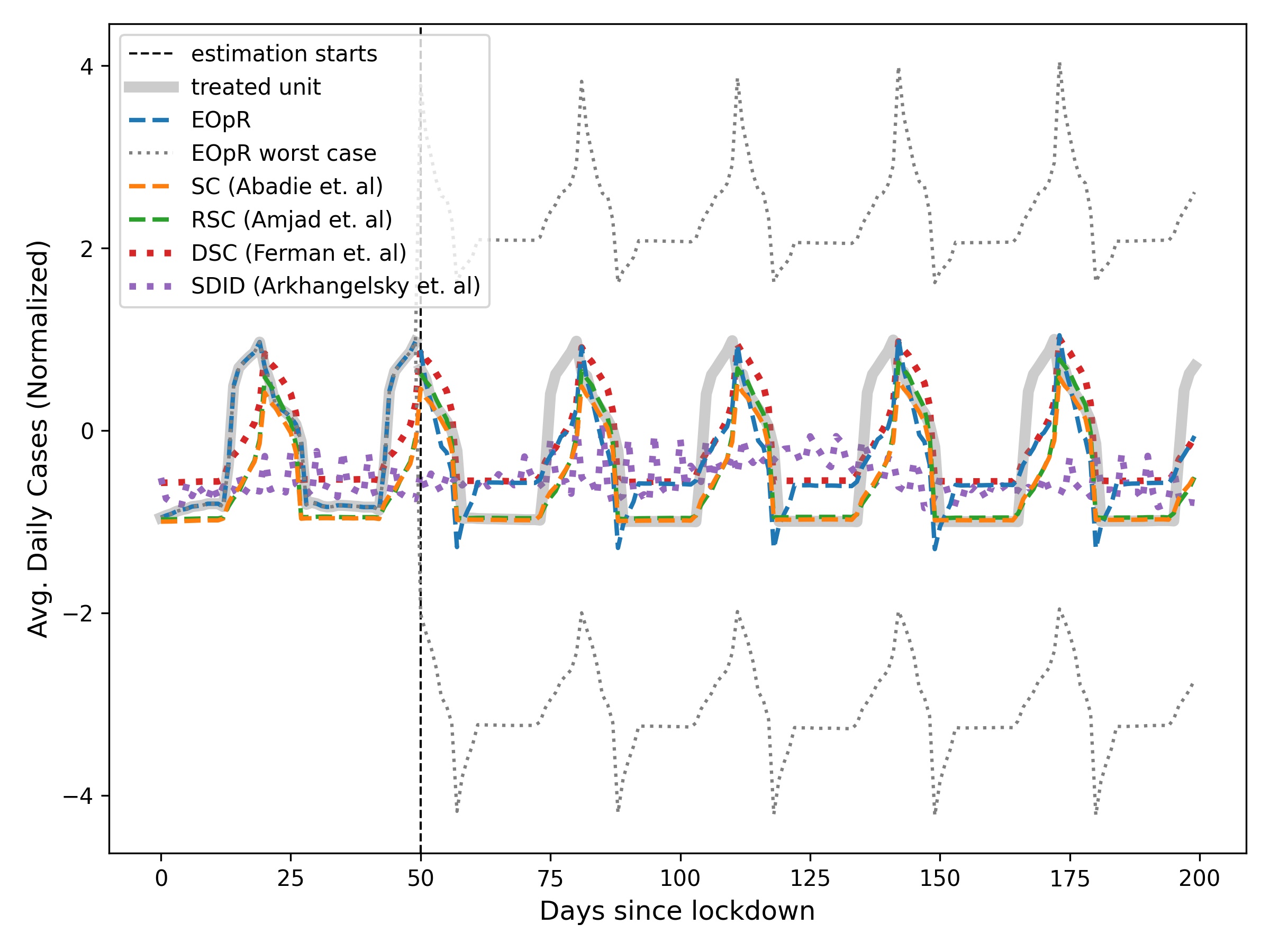}}
\vfill
    \subfloat[$N=7$]{
    \includegraphics[width=0.7\linewidth]{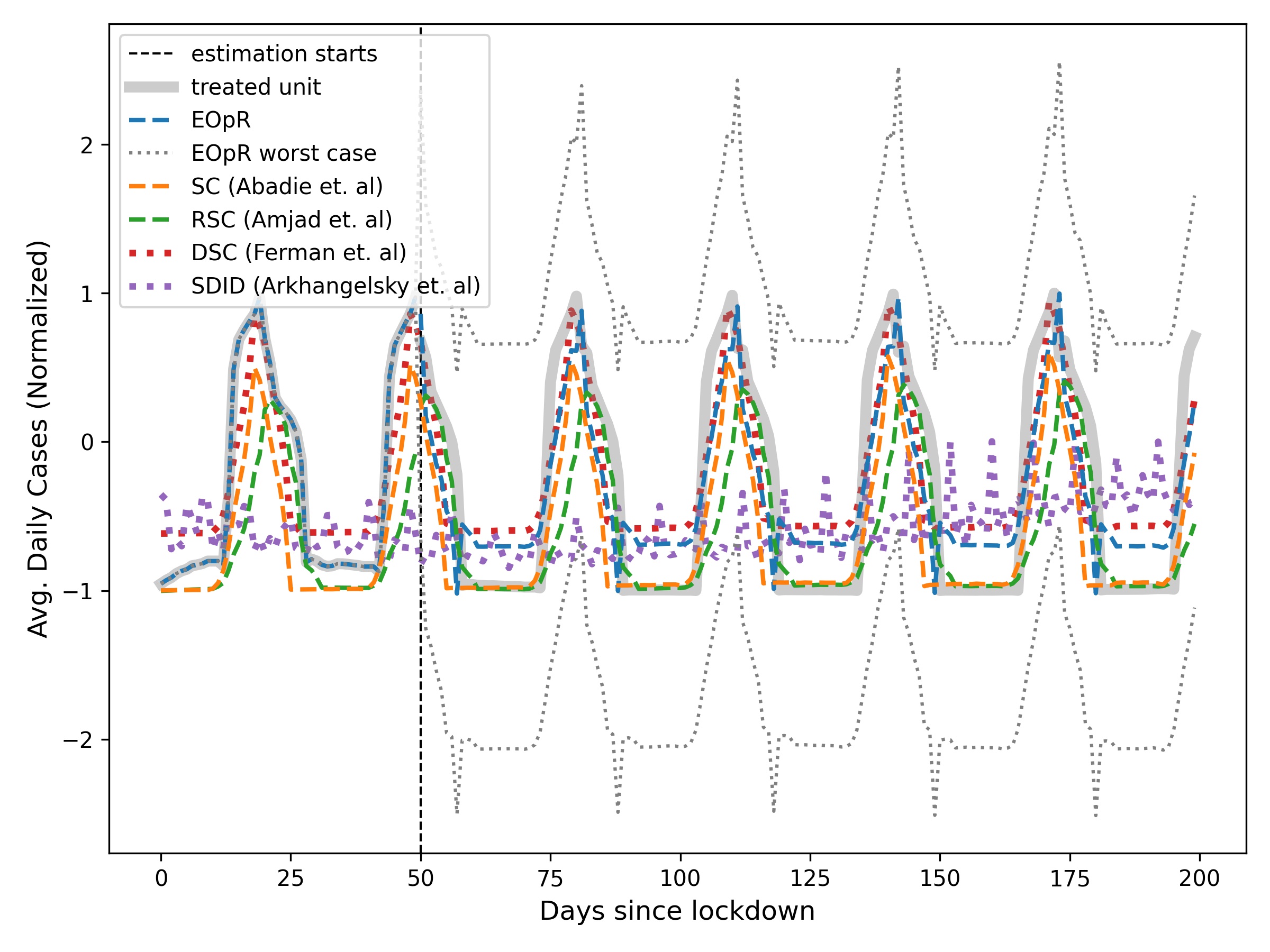}}
\vfill
    \subfloat[$N=15$]{
    \includegraphics[width=0.7\linewidth]{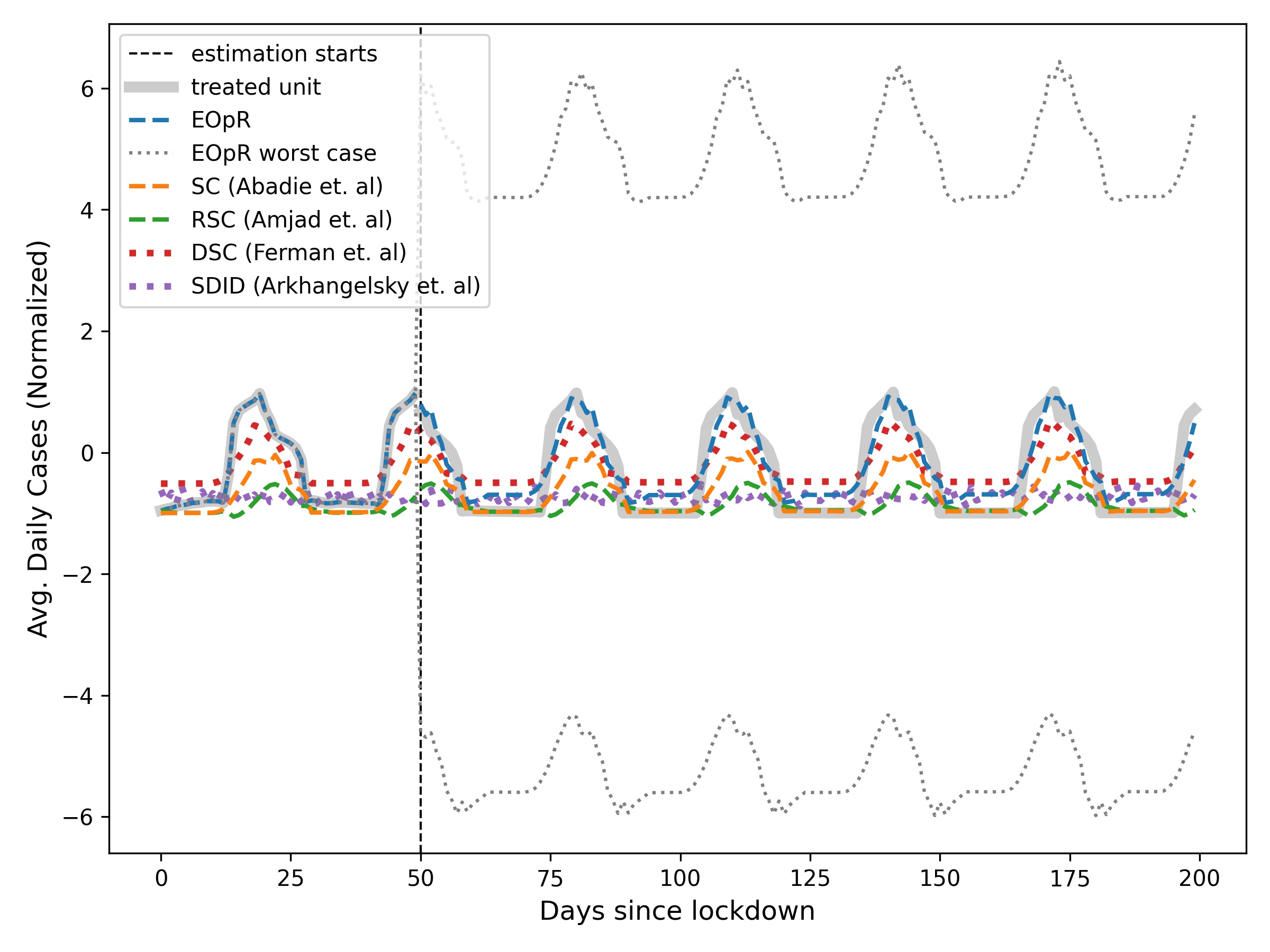}}
\vfill
    \subfloat[$N=43$]{
    \includegraphics[width=0.7\linewidth]{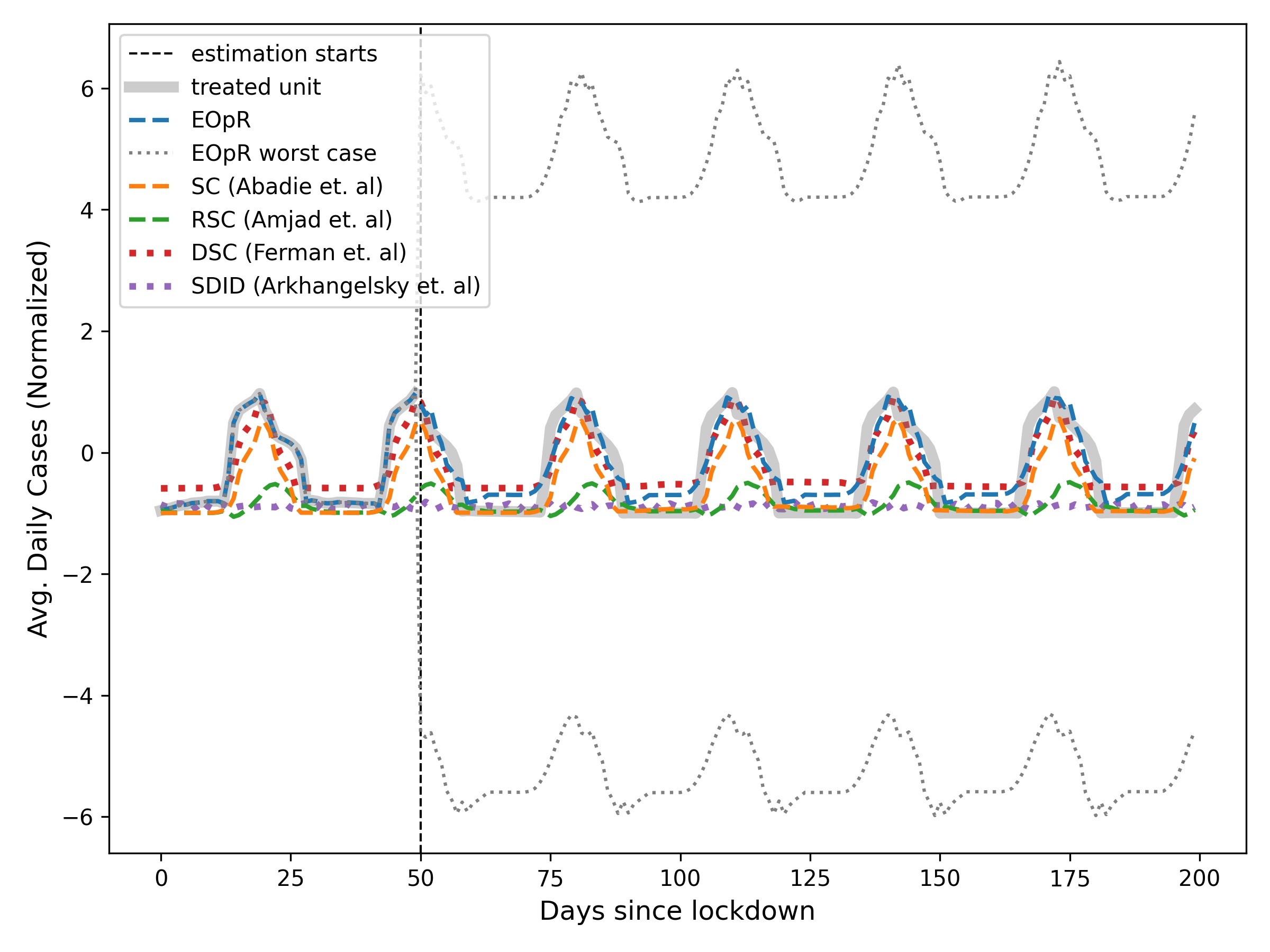}}

    \caption{Comparisons of methods for estimating New York trend}
    \label{fig:ny_estimation}
\end{figure}
Figure \ref{fig:ny_estimation} shows the estimated trend of daily average cases of New York, the estimation starts at $T_0$, in comparison to the actual New York trend (treated unit in the figure). The worst-case estimates are also plotted. Note that the worst-case estimates range changes depending on the quality of control units. 

\subsection{Ablation Study}
\begin{figure}[!ht]
    \centering
    \subfloat[Optimized $\lambda$\label{fig:lambda1}]{
\includegraphics[width=0.8\linewidth]{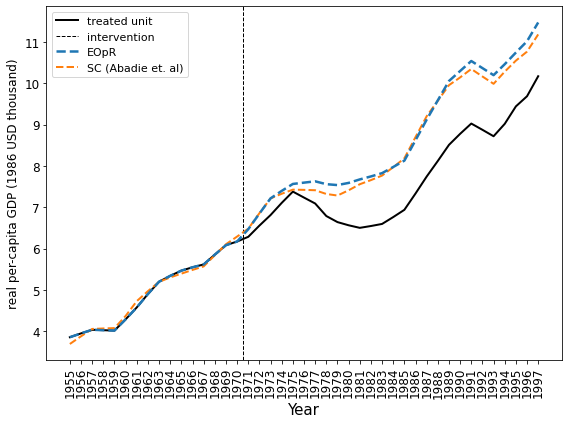}}
\vfill
    \subfloat[$\lambda = 0$\label{fig:lambda2}]{
    \includegraphics[width=0.8\linewidth]{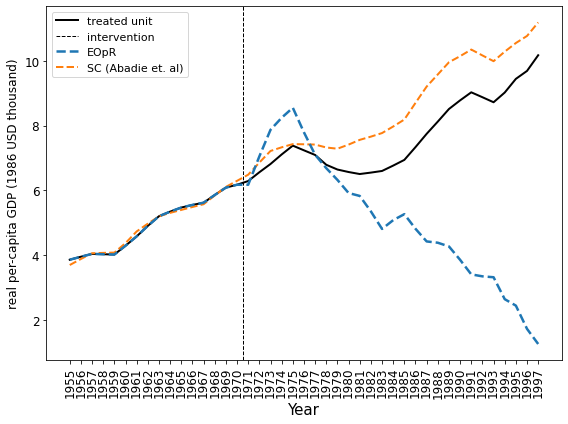}}

    \caption{Impact of adding $\lambda$ to the covariance matrix $\mat{S}\mat{S}^T$ for Basque case study}
    \label{fig:labmda_variations}
\end{figure}

We empirically study the effect of the added perturbation $\lambda$ to the covariate matrix $\mat{SS^{\top}}$. This small perturbation is added to ensure $Q$ is positive definite in \eqref{eq:Q}, following the definition of ellipsoids \ref{def:ellipsoid}. When $\lambda = 0$, we see the resulting predictions suffer drastically, deviating from expected possible outcomes (Figure \ref{fig:lambda2}). Using an appropriate $\lambda$ as in Figure \ref{fig:lambda2}, balances the model complexity which helps safeguard the algorithm from potentially underfitting the training data, and producing a biased estimate. 

\begin{figure}[!ht]
    \centering
    \subfloat[pre-intervention error]{
\includegraphics[width=0.8\linewidth]{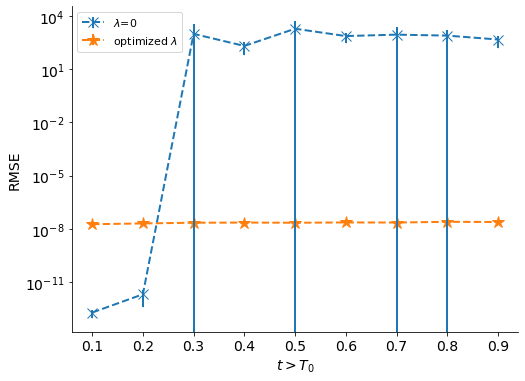}}
\vfill
    \subfloat[post-intervention error]{
    \includegraphics[width=0.8\linewidth]{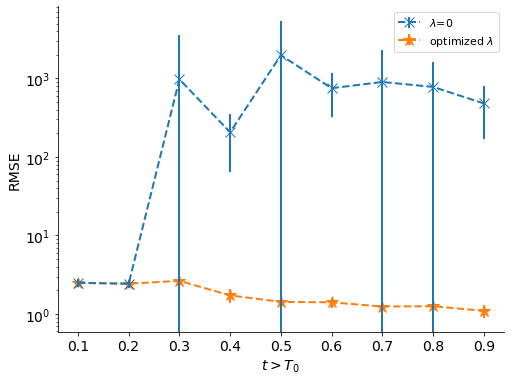}}

    \caption{Impact of adding $\lambda$ to the covariance matrix $\mat{S}\mat{S}^T$ for Basque case study}
    \label{fig:abalation_synthetic}
\end{figure}

A similar observation is seen for simulated data, for the setting of fixed $N$ and $T$ but changing the time of intervention $T_0$, in Figure \ref{fig:abalation_synthetic}.

    







\section{Conclusion}
Classical synthetic control has been noted as effective for causal inference in comparative studies. Here, we propose a signal processing approach for synthetic control---ellipsoidal optimal recovery (EOpR)---that estimates the causal effect given a policy intervention. Further, given the properties of EOpR, we derive worst-case estimates, which are themselves very useful for policy evaluation. Our approach of EOpR has less estimation error for pre- and post-intervention periods, especially with short pre-intervention periods. This is demonstrated through comparisons on both simulated data and classical case studies in econometrics. Applications in health-relevant settings are also shown to be compelling. Placebo tests and ablation studies demonstrate robustness. Extensions to the basic optimal recovery framework beyond ellipsoidal signal classes that we develop here are also possible.

\section*{Acknowledgement}
Authors would like to thank Akhil Bhimaraju and Alayt Issak for their useful insights and thoughtful discussions.  

\bibliographystyle{IEEEtran} 
\bibliography{ibtihal_ref}

\appendices

\input{apx.tex}

\end{document}

%% file: tables/covid_movingAVG.tex
\begin{table}[!ht]
    \centering
    \caption{Pre- and Post-intervention RMSE for COVID-19 case study}
    \scalebox{0.83}{
      \begin{tabular}{c|cccc}
      \toprule
      \multicolumn{1}{c}{} & \multicolumn{4}{c}{Pre-intervention error} \\
      \midrule
      $N$ & 5   & 7   & 15  & 43 \\
      \midrule
      EOpR & \textbf{1.9$\times 10^{-14}$} & \textbf{1.8$\times 10^{-14}$} & \textbf{3.2$\times 10^{-14}$} & \textbf{1.3$\times 10^{-13}$} \\
      SC (Abadie et al.) & 0.58 & 0.5 & 0.63 & 0.52 \\
      RSC (Amjad et al.) & 0.65 & 0.63 & 0.91 & 0.86 \\
      DSC (Ferman et al.) & 0.41 & 0.32 & 0.42 & 0.33 \\
  SDID (Arkhangelsky et. al) & 0.80  & 0.80  & 0.86  & 0.94 \\
      \midrule
          & \multicolumn{4}{c}{Post-intervention error} \\
      \midrule
      $N$ & 5   & 7   & 15  & 43 \\
      \midrule
      EOpR & 0.5 & \underline{0.4} & \textbf{0.3} & \textbf{0.3} \\
      SC (Abadie et al.) & 0.54 & 0.5 & 0.57 & 0.49 \\
      RSC (Amjad et al.) & 0.54 & 0.52 & 0.85 & 0.8 \\
      DSC (Ferman et al.) & \textbf{0.44} & \underline{0.4} & 0.47 & 0.41 \\
  SDID (Arkhangelsky et. al) & 0.78  & 0.81 & 0.85  & 0.93 \\
      \bottomrule
      \end{tabular}%
      }
    \label{tab:covid_movingAVG_errors}%
  \end{table}%

%% file: apx.tex
\section{Consistency and Unbiasedness}
\subsection{Analytical Proof}
\label{apx:analytical_proof}
To show consistency, here we bound the $\ell_2$ error of the estimation. We will drop the dependency on $\lambda$. Recall that the noise term $\epsilon_1$ is a zero-mean independent random variable that satisfies $\mathbb{E}(\epsilon_{ij}) = 0$ for all $i$ and $j$ by assumption, with variance $Var(\epsilon_{ij}) = \sigma^2$. 

\begin{lemma}
Suppose $x_1 = s_1+\epsilon_1$ with $\mathbb{E}(\epsilon_{1j}) = 0$ and $Var(\epsilon_{1j}) \leq \sigma^{2}$ for all $j \in \{1, \dots, T\}$.  Let $w^{\ast}$ be the optimal weights as in \eqref{eq:weights_extrapolation}, and $\hat{w}$ be the sub-optimal minimizer. Then for any $\lambda \geq 0$,
\begin{equation}
    \mathbb{E}\lVert s_1 - \hat{s_1}\rVert \leq 2 \sigma^2|r| \mbox{.}
\end{equation}
\end{lemma}
\textit{Proof.} Recall from \eqref{eq:factor_model} that the treatment row $x_1 = s_1+\epsilon_1$. By the definition of the Chebyshev center and its properties of unique estimation, consider $s_1 = \Sigma^T w^{\ast}$ with exact recovery, and $\hat{s_1} = \Sigma^T\hat{w}$. 
\begin{equation} 
    \begin{split}
        \lVert s_1 - \hat{s}_1 \rVert^2 & = \lVert (x_1 - \epsilon_1) - \Sigma^T \hat{w} \lVert^2 \\
        & = \lVert (x_1 - \Sigma^T \hat{w} + (-\epsilon_1) \rVert^2 \\
        & = \lVert x_1 - \Sigma^T\hat{w} \rVert^2 + \lVert \epsilon \rVert^2 + 2 \langle -\epsilon, x_1 - \Sigma^T \hat{w} \rangle \\
        & \leq \lVert x_1 - \Sigma^T w^{\ast} \rVert^2 
        + \lVert \epsilon_1 \rVert^2 
        + 2 \langle -\epsilon, x_1 - \Sigma^T \hat{w} \rangle \\
        & = \lVert (\Sigma^T w^{\ast} + \epsilon_1) - \Sigma^T w^{\ast}\rVert^2 
        + \lVert \epsilon_1 \rVert^2 
        \\  & \quad + 2 \langle -\epsilon, x_1 - \Sigma^T \hat{w} \rangle \\
        & = 2 \lVert \epsilon_1 \rVert^2 + \rVert^2 
        + 2 \langle -\epsilon, x_1 - \hat{\Sigma}^T \hat{w} \rangle \mbox{.}
    \end{split}
\end{equation}
Taking expectations, we arrive at the inequality 
\begin{equation}
    \mathbb{E} \lVert s_1 - \hat{s}_1 \rVert \leq 
    2 \mathbb{E} \lVert \epsilon_1 \rVert^2 
    + 2 \mathbb{E} \langle -\epsilon, x_1 - \hat{\Sigma}^T \hat{w} \rangle \mbox{.}
\label{eq:expectation_inequality}
\end{equation}
We must address an inner product on the right side. First, we derive some useful facts. Recall the trace operator has the mapping property $\mathit{tr}(AB) = \mathit{tr}(BA)$, and the projection matrix $P$ to be $P_1 = AA^{\dag}$ and $P_2 = A^{\dag}A$. Hence, 
\begin{equation}
\begin{split}
    \mathbb{E}[(\epsilon_1)^T\Sigma^T\Sigma^{{\dag}^T}\epsilon_1] 
    & =  \mathbb{E}[tr((\epsilon_1)^T\Sigma^T\Sigma^{\dag}\epsilon_1)] \\
    & = \mathbb{E}[tr(\Sigma^T\Sigma^{\dag})\epsilon_1 (\epsilon_1)^T] \\
    & = \mathit{tr} \big( \mathbb{E} [\Sigma^T \Sigma^{\dag^T} \epsilon_1 (\epsilon_1)^T] \big) \\
    & = \mathit{tr} \big( \mathbb{E} [\Sigma^T \Sigma^{\dag^T}] \mathbb{E} [\epsilon_1 (\epsilon_1)^T] \big) \\
    & = \mathit{tr}\big(\mathbb{E}[\Sigma^T\Sigma^{{\dag}^T}]\sigma^2 I\big) \\
    & = \sigma^2 \mathbb{E}[tr(\Sigma^T\Sigma^{{\dag}^T})] \\
    & = \sigma^2 \mathbb{E}[\text{rank}(\Sigma)] \\
    & \leq \sigma^2 |r| \mbox{,}
\end{split}
\end{equation}
which follows since the trace of a projection matrix equals the rank of the matrix, i.e.\ $tr(\Sigma^T \Sigma^{\dag^T}) = \text{rank}(\Sigma^T)$. Hence, the rank of $\Sigma$ is at most $|r|$. 

Returning to the inner product, recall $\hat{w} = \Sigma^{{\dag}^T} s_1$ from \eqref{eq:s_extrapolation}. 
\begin{equation}
    \begin{split}
        \mathbb{E}[\langle \epsilon_1, x_1 - \Sigma\hat{w}\rangle] 
        & = \mathbb{E}[(\epsilon_1)^T\Sigma \hat{w}]-\mathbb{E}[(\epsilon_1)^Tx_1] \\
        & = \mathbb{E}[(\epsilon_1)^T\Sigma^T \Sigma^{\dag^T}x_1] -\mathbb{E}[(\epsilon_1)]s_1 - \mathbb{E}[(\epsilon_1)^T\epsilon_1] \\ 
        & = \mathbb{E}[(\epsilon_1)^T\Sigma^T\Sigma^{\dag^T}]s_1 
        + \mathbb{E}[(\epsilon_1)^T\Sigma^T\Sigma^{\dag^T} \epsilon_1] 
        -  \mathbb{E}[(\epsilon_1)^T\epsilon_1] \\
        & = \mathbb{E}[(\epsilon)^T][\Sigma^T\Sigma^{\dag^T}]s_1 
        + \mathbb{E}[(\epsilon_1)^T\Sigma^T\Sigma^{\dag^T} \epsilon_1] 
        -  \mathbb{E}[(\epsilon_1)^T\epsilon_1] \\
        & = \mathbb{E}[(\epsilon_1)^T\Sigma^T\Sigma^{\dag^T} \epsilon_1] 
        -  \mathbb{E}\lVert \epsilon_1 \rVert^2 \\
        & \leq \sigma^2|r| -  \mathbb{E}\lVert \epsilon_1 \rVert^2 \mbox{.}
    \end{split}
\end{equation}
Finally, we replace the above terms into inequality \eqref{eq:expectation_inequality} to arrive at
\begin{equation}
\begin{split}
    \mathbb{E}\lVert s_1 - \hat{s_1} \rVert^2 
    & \leq 2 \mathbb{E} \lVert \epsilon_1 \rVert^2 
    + 2 \mathbb{E} \langle -\epsilon, x_1 - \hat{\Sigma}^T \hat{w} \rangle \\
    & \leq 2 \sigma^2|r| \mbox{,}
\end{split}
\end{equation}
which completes the proof. 

Now, let us consider replacing the above value in the mean squared error ($\mathit{MSE}$) function as
\begin{equation}
\begin{split}
    \mathit{MSE}(s_1, \hat{s}_1) & = \frac{1}{T}\lVert s_1 - \hat{s}_1 \rVert^2  \\
    & \leq \frac{2 \sigma^2|r|}{T} \mbox{.}
\end{split}
\end{equation}

\subsection{Geometric proof}
\label{apx:geometric_proof}
The following theorem \cite{halteman1986chebyshev} demonstrates that the Chebyshev center is an unbiased and consistent estimator for the center of a normal distribution over a $k$-sphere. The result can further be extended from a sphere  to an ellipse. 

\begin{theorem}[Unbiasedness and consistency] Suppose the points $W = \{w_i, \dots, w_n\}$ are sampled from an independent and identically distributed (i.i.d.) spherical $k$-dimensional distribution to create a sphere $S(\mu, r)$, with center $\mu$ and radius $r$. Let $S(\bar{\mu}, \bar{r})$, be the smallest sphere containing the samples of $W$, (i.e. $\bar{r} < r$). The center $\bar{\mu}$, a Chebyshev center, has a spherical distribution and hence $\bar{\mu}$ is unbiased for $\mu$, and consistent, i.e. $\bar{\mu} \rightarrow \mu$ with probability goes to unity. 
\end{theorem}
\textit{Proof.} Given that distribution of $S(\mu, r)$ is i.i.d., it is invariant under any rotation about $\mu$. The center $\bar{\mu}$, for any rotated sample in $W$, will be rotated by an equal amount of $\mu$. Therefore, the distribution of $\bar{\mu}$ is rotationally invariant about the same $\mu$, and thus the Chebyshev center $\bar{\mu}$ is unbiased. 

To prove consistency, let the sphere $S(\mu, r)$ have density function $f(w)=\frac{1}{n}$, and let $H$ be the set of points at the boundary of $S(\bar{\mu}, \bar{r})$. Then, the maximum distance $d$ between $w\in H$ and the sphere $S(\mu, r)$ converges to $0$ for large $n$, and so  $\bar{\mu} \rightarrow \mu$ with probability $1$. 